\documentclass{jfm}
\usepackage{graphicx}
\usepackage{epsfig} 
\usepackage{newtxtext}
\usepackage{newtxmath}
\graphicspath{{Figures/}}
\usepackage{multirow}
\usepackage{mathrsfs,amsmath}
\usepackage{bm}
\usepackage{amssymb}
\usepackage{multirow,bigdelim}
\usepackage{upgreek}
\usepackage{units}
\usepackage{array,multirow}
\usepackage{booktabs}
\usepackage{psfrag}
\usepackage{tikz}
\usetikzlibrary{arrows.meta}
\usepackage{color}
\usepackage{threeparttable}	
\usepackage{dcolumn}		
\newcolumntype{d}[1]{D{.}{.}{#1}}
\usetikzlibrary{decorations}
\usetikzlibrary{decorations.markings}
\usetikzlibrary{calc,decorations.pathreplacing}
\usepackage{lipsum}
\usepackage{mathtools}
\usepackage{xfrac}
\usepackage{ar}
\usepackage{enumitem}
\usepackage{url}
\usepackage{subcaption}
\usepackage{float}
\usepackage{tabularx}
\usepackage{natbib}
\usepackage{hyperref}
\hypersetup{
    colorlinks = true,
    linkcolor = {blue},
    citecolor  = {black},
    filecolor = {blue},
    menucolor= {blue},
    runcolor = {blue},
    urlcolor = {blue},
}

\newcommand{\RomanNumeralCaps}[1]
\linenumbers

\shorttitle{Wall-pressure--velocity correlations in wall-bounded turbulence}
\shortauthor{Baars, Dacome, and Lee}

\title{Reynolds-number scaling of wall-pressure--velocity correlations in wall-bounded turbulence}

\author{Woutijn J. Baars\aff{1}\dag,
  Giulio Dacome\aff{1}
 \and Myoungkyu Lee\aff{2}\corresp{\email{w.j.baars@tudelft.nl, leemk@uh.edu}}}
 
\affiliation{\aff{1}Faculty of Aerospace Engineering, Delft University of Technology, 2629 HS Delft, The Netherlands
\aff{2}Department of Mechanical Engineering, University of Houston, Houston, TX 77204, U.S.A.}
\begin{document}

\maketitle
\begin{abstract}
Wall-pressure fluctuations are a practically robust input for real-time control systems aimed at modifying wall-bounded turbulence. The scaling behaviour of the wall-pressure—velocity coupling requires investigation to properly design a controller with such input data so that it can actuate upon the desired turbulent structures. A comprehensive database from direct numerical simulations of turbulent channel flow is used for this purpose, spanning a Reynolds-number range $Re_\tau \approx 550$ -- $5\,200$. Spectral analysis reveals that the streamwise velocity is most strongly coupled to the linear term of the wall-pressure, at a Reynolds-number invariant distance-from-the-wall scaling of $\lambda_x/y \approx 14$ (and $\lambda_x/y \approx 8$ for the wall-normal velocity). When extending the analysis to both homogeneous directions in $x$ and $y$, the peak-coherence is centred at $\lambda_x/\lambda_z \approx 2$ and $\lambda_x/\lambda_z \approx 1$ for $p_w$ and $u$, and $p_w$ and $v$, respectively. A stronger coherence is retrieved when the quadratic term of the wall-pressure is concerned, but there is only little evidence for a wall-attached-eddy type of scaling. An experimental dataset comprising simultaneous measurements of wall-pressure and velocity complements the DNS-based findings at one value of $Re_\tau \approx 2$k, with ample evidence that the DNS-inferred correlations can be replicated with experimental pressure data subject to significant levels of (acoustic) facility noise. It is furthermore shown that velocity-state estimations can be achieved with good accuracy by including both the linear and quadratic terms of the wall-pressure. An accuracy of up to 72\,\% in the binary state of the streamwise velocity fluctuations in the logarithmic region is achieved; this corresponds to a correlation coefficient of $\approx 0.6$. This thus demonstrates that wall-pressure sensing for velocity-state estimation---\emph{e.g.,} for use in real-time control of wall-bounded turbulence---has merit in terms of its realization at a range of Reynolds numbers.
\end{abstract}

\section{Introduction}\label{sec:intro}
Inspiration for this work was born out of practical considerations associated with the implementation of real-time flow control for the reduction of skin-friction drag in wall-bounded turbulence. Efforts in turbulence control comprise both passive and active methods to target near-wall structures that scale in viscous units, $l^* \equiv \nu/u_\tau$, where $u_\tau \equiv \sqrt{\tau_w/\rho}$ is the friction velocity and $\rho$ and $\nu$ are the fluid's density kinematic viscosity, respectively. Leaving passive techniques aside, most studies on active skin-friction control tailor (statistical) forcing techniques to the \emph{inner scales} \citep[not requiring any sensing, \emph{e.g.},][]{choi:1998a,kasagi:2009a,choi:2011a,bai:2014a}; we refer to this as `predetermined forcing-control.' Only a few studies do incorporate \emph{sensing} in numerical \citep{choi:1994a,lee:1998a} and experimental \citep{rathnasingham:2003a,qiao:2018a} `real-time control' efforts of the near-wall structures. This approach requires that sensors and actuators are sized to said near-wall structures, which are roughly $500l^*$ in length and $100l^*$ in width. On practical engineering systems such as on an aircraft in cruise, $Re_\tau \approx 80$k at a typical location for control on the fuselage (here $Re_\tau \equiv \delta u_\tau/\nu$, with $\delta$ being the boundary layer thickness). As such, control of the near-wall scales requires sensors/actuators with a spatial scale of $\sim 0.1$\,mm (\emph{e.g.}, the size of sensors/actuators) and with a temporal scale in excess of 30\,kHz (\emph{e.g.}, the frequency of actuation). This required spatial-temporal resolution of sensors and actuators is out-of-reach for current technologies. In addition, even if near-wall scales can be sensed and partially disrupted, the flow is expected to recover to the uncontrolled state within a streamwise distance that scales in viscous units (recall the auto-regeneration mechanisms of near-wall turbulence \citep{hamilton:1995a}, and see the study by \citet{qiao:2018a} in which the controlled skin-friction drag recovers in less than 100 viscous units). Hence, the number of control-stations for achieving streamwise-persistent control is also impractical. To overcome these issues, alternative control strategies emerged in tandem with studies of higher-Reynolds number wall-turbulence. A direct numerical simulation (DNS) study by \citet{schoppa:1998a} utilized spanwise jets for predetermined, large-scale forcing-control and reported drag reductions of up to 50\%. However, the low Reynolds number of $Re_\tau = 180$ yielded a negligible large-scale energy content in terms of the bulk turbulence kinetic energy (TKE), and recent investigations debate the effectiveness at higher $Re_\tau$ \citep{canton:2016a,deng:2016a,yao:2018a}. With higher-Reynolds number studies available to date, experimental efforts to control flow structures within the logarithmic region in real-time were undertaken \citep{abbassi:2017a}. Even though the mean friction was favorably affected, the net energy savings of such control systems are difficult to assess and generalize.

Independent of the implementation or large-scale control, the theoretical work of \citet{renard:2016a} outlines the potential of this pathway in an elegant manner: their kinetic-energy budget analysis reveals how the skin-friction drag relates to different physical phenomena across the boundary layer. Due to the decay of the relative contributions of the buffer and wake regions to the TKE production with increasing $Re_\tau$ \citep{smits:2011a}, the generation of the turbulence-induced excess friction is dominated by the dynamics in the logarithmic region. Hence, the work of \citet{renard:2016a} suggests that drag reduction strategies targeting the TKE production mechanisms in that layer are worth investigating.
\begin{figure} 
\vspace{0pt}
\centering
\includegraphics[width = 0.999\textwidth]{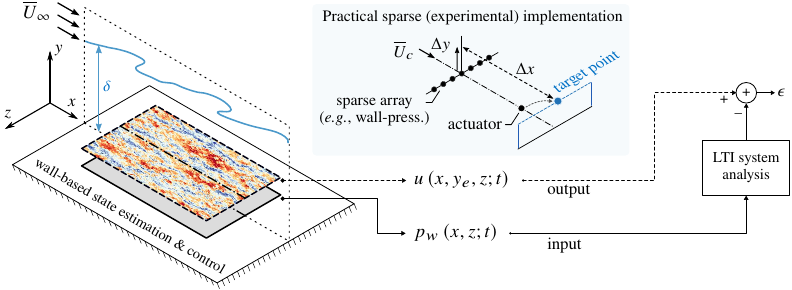}
\caption{Wall-based quantities are to be used for a linear time invariant (LTI) system analysis, to estimate the state of the off-the-wall turbulent velocities. A sparse implementation considers a limited number of sensors/actuators, and includes typical offsets in the wall-normal ($\Delta y$) and streamwise ($\Delta x$) directions between the sensing location and the controller's `target point'.}
\label{fig:ischem}
\end{figure}

A controller's ability to \emph{selectively target} certain turbulent structures depends on the degree of coupling between off-the-wall flow quantities within the linear, buffer, and/or logarithmic regions of the flow, and observable wall-quantities (sensors should be flush within the wall to avoid parasitic drag). When working towards realistic setups, a real-time control system with discrete sensors and actuators is the most logical (components should still be wall-embedded to minimize parasitic drag, but the entire wall itself is not used for conducting the control action). In addition, when bringing in the requirement to (selectively) manipulate structures \citep[\emph{e.g.},][]{abbassi:2017a}, a controller has to operate with an unavoidable wall-normal separation, $\Delta y$, between the sensing location and the `target point' (figure~\ref{fig:ischem}). Likewise, a separation in the streamwise direction, $\Delta x$, must be adopted to account for control latency. These spatial separations result in a loss-of-correlation between the controller's input and the grazing turbulent velocities. Therefore, a sound understanding of (the scaling of) wall-pressure--velocity correlations is a pre-requisite for designing a successful control system based on wall-pressure input data.

The notation in this paper is as follows. Coordinates $x$, $y$ and $z$ denote the streamwise, wall-normal, and spanwise directions of the flow, and $u$, $v$, $w$ and $p$ represent the Reynolds decomposed fluctuations of the three velocity components and the static pressure, respectively. The wall-pressure is denoted as $p_w$. Overlined capital symbols, \emph{e.g.}, $\overline{U}$, are used for the absolute mean. When quantities are presented in outer-scaling, length scale $\delta$ and velocity scale $\overline{U}_\infty$ are used, while a viscous scaling is signified with superscript `$+$' and comprises a scaling with length scale $l^* = \nu/u_\tau$ and velocity scale $u_\tau$.

\subsection{Wall-pressure--velocity correlations}\label{sec:introscaling}
When attempting large-scale control of wall-bounded turbulence based on wall-pressure input data, a first step is to estimate a \emph{dynamic state} of logarithmic region-turbulence from said fluctuating wall-pressure, $p_w$. The velocity fluctuations themselves form the \emph{true state}. Many studies are concerned with scaling laws and modeling attempts of pressure fluctuations \citep[\emph{e.g.},][]{willmarth:1975a,farabee:1991a,klewicki:2008a,hwang:2009a}, but only very few works are concerned with an assessment of the instantaneous or statistical coupling between the velocity fluctuations and the wall-pressure. \citet{thomas:1983a} revealed characteristic wall-pressure signatures associated with the near-wall burst-sweep events. Throughout the last decade, high-resolution mappings of the spatio-temporal pressure-velocity correlation have been reported \citep{ghaemi:2013a,naka:2015a}, but for a specific $Re_\tau$ condition. More recently, input-output linear time-invariant (LTI) system analyses have been conducted in the experimental works of \citet{blitterswyk:2017a} and \citet{gibeau:2021a}. Nevertheless, no studies exist to date addressing the scaling of input-output relations, as a function of $y$ and $Re_\tau$. It is challenging to simultaneously acquire velocity and (noise-free) pressure data, particularly in wall-parallel planes as shown in figure~\ref{fig:ischem}, while such data are needed for examining wall-pressure--velocity correlations as a function of the streamwise and spanwise wavenumbers. Novel experiments can be designed to yield 2D coherence spectra (see \citet{deshpande:2020a} for velocity-velocity correlations), but these are not trivial to perform. Given the current knowledge gap on wall-pressure--velocity coupling, our work covers an LTI-system analysis using high-fidelity data available from DNS campaigns.

A model transfer kernel between wall-pressure and velocity is particularly useful for wall-based estimations of turbulent velocities. Upon confining the analysis to stochastic estimation-based techniques, these have proven to be useful for estimating large-scale features in turbulent flows using sparse input data. First-order techniques, known as Linear Stochastic Estimation (LSE), were initially introduced in the turbulence community to inspect coherent turbulent structures in shear flows \citep{adrian:1979a,adrian:1988a}. LSE can be implemented following a single- or multi-offset approach \citep{ewing:1999c}, in which the latter is identical to a spectral approach \citep[see][]{tinney:2006a}. The transfer kernel of the spectral LSE approach accounts for a gain and offset (or phase) per temporal and/or spatial scale, depending on the implementation. In view of figure~\ref{fig:ischem}, an LSE of an off-the-wall velocity field $u(x,y_e,z)$ at an estimation position $y_e$ starts with a 2D spatial Fourier transform of the unconditional input field $p_w(x,z)$,
\begin{equation}\label{eq:introLSE1}
    P_w\left(\lambda_x,\lambda_z\right) = \mathcal{F}\left[p_w(x,z)\right].
\end{equation}
A spectral-domain estimate is then formulated as
\begin{equation}\label{eq:introLSE2}
    \widehat{U}_{\rm LSE}\left(\lambda_x,\lambda_z,y_e\right) = H_L\left(\lambda_x,\lambda_z,y_e\right)P_w\left(\lambda_x,\lambda_z\right),
\end{equation}
and the physical-domain estimate is found through the inverse Fourier transform,
\begin{equation}\label{eq:introLSE3}
    \widehat{u}_{\rm LSE}(x,y_e,z) = \mathcal{F}^{-1}\left[\widehat{U}_{\rm LSE}\left(\lambda_x,\lambda_z,y_e\right) \right].
\end{equation}
The stochastic and complex-valued linear kernel in \eqref{eq:introLSE2} is computed \emph{a priori} and is equal to the wall-pressure--velocity cross-spectrum, $\phi_{up_w}$, divided by the wall-pressure input spectrum, $\phi_{p_wp_w}$, following
\begin{equation}\label{eq:introHL}
    H_L\left(\lambda_x,\lambda_z,y_e\right) = \frac{\phi_{up_w}\left(\lambda_x,\lambda_z,y_e\right)}{\phi_{p_wp_w}\left(\lambda_x,\lambda_z\right)}.
\end{equation}
Throughout this paper a spectrum is defined as $\phi_{up_w}\left(\lambda_x,\lambda_z,y_e\right) = \langle U\left(\lambda_x,\lambda_z,y_e\right)P^*_w\left(\lambda_x,\lambda_z\right)\rangle$; capital quantities refer to the Fourier transform, $U = \mathcal{F}\left[u\right]$, the asterisk $*$ indicates the complex conjugate, and $\langle\cdot\rangle$ denotes ensemble-averaging.

To deduce how much energy of the LTI-system's output can be estimated, the linear coherence spectrum (LCS) between the input and output data is insightful, and is defined as
\begin{equation}\label{eq:introLCS}
    \gamma^2_{up_w}\left(\lambda_x,\lambda_z,y_e\right) = \frac{\vert\phi_{up_w}\left(\lambda_x,\lambda_z,y_e\right)\vert^2}{\phi_{uu}\left(\lambda_x,\lambda_z,y_e\right)\phi_{p_wp_w}\left(\lambda_x,\lambda_z\right)} \in [0,1].
\end{equation}
Using \eqref{eq:introHL} the coherence can be rewritten as $\gamma^2_{up_w} = \vert H_L \vert^2\phi_{p_wp_w}/\phi_{uu}$. In a stochastic sense, the coherence magnitude is thus interpreted as the energy in the estimated output signal ($\vert H_L \vert^2\phi_{p_wp_w}$), relative to the true output energy ($\phi_{uu}$). An assessment of the coherence between the wall-pressure and the turbulent velocities will address whether a substantial amount of energy in the turbulent fluctuations can be estimated. 

So far, only the linear wall-pressure term was considered and this confines the analysis to scale-by-scale interactions. LSE must be extended to a Quadratic Stochastic Estimation (QSE) when nonlinearities manifest themselves in the input-output relation. \citet{naguib:2001a} methodized a time-domain QSE and showed that the quadratic terms are critical for satisfactory estimates of the conditional streamwise velocity based on wall-pressure events in turbulent boundary layer (TBL) flows. Another example of improved estimates with QSE over LSE includes the estimate of velocities in a cavity shear layer, based on wall-pressure \citep{murray:2003a,murray:2004c,lasagna:2013a}. By including the second-order, quadratic term in the stochastic estimate, a QSE procedure for the off-the-wall velocity field is formulated as
\begin{equation}\label{eq:introQSE}
    \widehat{U}_{\rm QSE}\left(\lambda_x,\lambda_z,y_e\right) = \underbrace{H^\prime_L\left(\lambda_x,\lambda_z,y_e\right)P_w\left(\lambda_x,\lambda_z\right)}_{\text{linear term}} + \underbrace{H_Q\left(\lambda_x,\lambda_z,y_e\right)P_{w^2}\left(\lambda_x,\lambda_z\right)}_{\text{quadratic term}}.
\end{equation}
When the skewness of the wall-pressure is zero, it can be shown that the linear kernel in \eqref{eq:introQSE} is identical to \eqref{eq:introHL}, thus $H^\prime_L = H_L$. For details, we refer to \citet{naguib:2001a} and reported observations of negligible wall-pressure skewness \citep{gravante:1998a,tsuji:2007a,klewicki:2008a}. Regarding the quadratic term, this one consists of the Fourier transform of the quadratic wall-pressure, $P_{w^2}\left(\lambda_x,\lambda_z\right) = \mathcal{F}\left[p_w^2(x,z)\right]$. The pressure-squared term $p_w^2$ is computed as the square of the de-meaned wall-pressure field. The quadratic kernel $H_Q\left(\lambda_x,\lambda_z,y_e\right)$ is computed based on the same wall-pressure-squared term, following \eqref{eq:introHQ}, and under the condition that the wall-pressure skewness is negligible.
\begin{equation}\label{eq:introHQ}
    H_Q\left(\lambda_x,\lambda_z,y_e\right) = \frac{\phi_{up_w^2}\left(\lambda_x,\lambda_z,y_e\right)}{\phi_{p_w^2p_w^2}\left(\lambda_x,\lambda_z\right)}.
\end{equation}

\citet{naguib:2001a} attributed the significant improvement of their conditional estimates of streamwise velocity fields based on wall-pressure events---with the inclusion of the quadratic pressure term---to the non-negligible turbulent--turbulent source (the `slow' nonlinear pressure source associated with large-scale motions). This was analysed by considering the wall-pressure dependence on the turbulent flow field, following the solution of Poisson's equation for incompressible flow. When estimates were based on wall-shear stress, quadratic terms were less crucial \citep{adrian:1987a,guezennec:1989a}. \citet{naguib:2001a} hypothesized that the portion in the estimate from the quadratic term represents a flow structure obeying outer scaling. However, they remained inconclusive due to their relatively small Reynolds number range. 

\subsection{Present contribution and outline}\label{sec:outline}
In summary, the extent of coupling between wall pressure and velocity at various friction Reynolds numbers in wall-bounded turbulence remains undetermined. Detailing the stochastic coupling, with linear and quadratic wall-pressure terms, is critical for gaining practical insight into whether a control system that relies on wall-pressure input has merit for real-time estimating (and controlling) large-scale structures in the logarithmic region. Albeit a wide variety of estimation techniques can be applied, such as physics-informed models \citep[\emph{e.g.}, based on linearized Navier-Stokes equations (NSE)][]{madhusudanan:2019a}, resolvent-based methods assimilating nonlinearity of the NSE \citep{arun:2023a} or neural networks \citep{guastoni:2021a}, the simplicity of the LTI-system analysis allows for interpretability.

This work is structured as follows. First, DNS data are presented in \S\,\ref{sec:datanum} and analysed through an input-output LTI-system approach in \S\,\ref{sec:yscaling} to infer Reynolds-number scaling relations for the wall-pressure--velocity correlations. Experimental wall-pressure and velocity data are also described (\S\,\ref{sec:dataexp}) and assessed to identify whether the correlations can be replicated with experimental pressure data that are subject to significant levels of (acoustic) facility noise. Subsequently, the accuracy of velocity-state estimations is covered in \S\,\ref{sec:state}, and this aspect of wall-pressure sensing for estimation is of high practical relevance when addressing whether control based on wall-pressure is a realistic route forward for real-time control.

\section{Numerical and experimental data}\label{sec:data}
\subsection{Direct numerical simulation of turbulent channel flow}\label{sec:datanum}
Four incompressible turbulent channel flow datasets are used and span one decade in Reynolds number, $Re_\tau \approx 550$ -- $5\,200$. Details on the numerical scheme, resolution, and turbulence statistics are documented by \citet{lee:2015a,lee:2019a}, and parameters of the DNS data and case names are summarized in table~\ref{tab:data}. In relevance to the current work involving pressure, the static pressure fields were obtained by solving the Poisson equation,
\begin{equation}\label{eq:peq1}
    \frac{\partial^2 p}{\partial x_k \partial x_k} = -\rho \frac{\partial^2 u_i u_j}{\partial x_i \partial x_j},
\end{equation}
based on the full velocity field at each instant snapshot. Neumann boundary conditions were applied at the walls, following
\begin{equation}\label{eq:peq2}
    \frac{\partial p}{\partial y} = \nu \frac{\partial^2 v}{\partial y^2}.
\end{equation}
Finally, the average pressure in the homogeneous directions was set according to,
\begin{equation}\label{eq:peq3}
    \langle p \rangle = - \rho \langle v^2 \rangle + C.
\end{equation}
Here, $\langle \cdot \rangle$ denotes the average in the homogeneous direction, and integration constant $C$ is set to zero for convenience. Later on, not only wall-pressure fields are used, but also fields of the wall-pressure squared $p_w^2$. Such nonlinear products during simulations and post-processing steps were generated using a $\nicefrac{3}{2}$ de-aliasing rule with zero-padding \citep{orszag:1971a}. Details of the pressure field computations are described in \citet{panton:2017a}.
\begin{table}
\begin{center}
\def~{\hphantom{0}}
\begin{tabular}{lllllll}
\multicolumn{6}{l}{\textbf{Numerical data:} Channel DNS of \citet{lee:2015a}} & \\\noalign{\smallskip}
~& Case: & R0550 & R1000 & R2000 & R5200 & ~\\
\noalign{\smallskip}\cline{2-6}\noalign{\smallskip}
~& $Re_\tau$ & 544 & 1\,000 & 1\,995 & 5\,186 & ~\\
~& $(L_x,L_z)/\delta$ & $(8\pi,3\pi)$ & $(8\pi,3\pi)$ & $(8\pi,3\pi)$ & $(8\pi,3\pi)$ & ~\\
~& $\Delta x^+$	& 8.9 & 10.9 & 12.2 & 12.7 & ~\\
~& $\Delta z^+$ & 5.0 & 4.6 & 6.1 & 6.4 & ~\\
~& $\Delta y_w^+,\Delta y_c^+$ & 0.019,\,4.5 & 0.019,\,6.2 & 0.017,\,8.2 & 0.498,\,10.3 & ~\\
~& $Tu_\tau/\delta$\footnotemark[1] & 13.6 & 12.5 & 11.5 & 7.80 & ~\\
~\\
\multicolumn{7}{l}{\textbf{Experimental data:} TBL at $x = 3.07$\,m in the W-tunnel} \\\noalign{\smallskip}
~& \multicolumn{2}{l}{Boundary layer} & \multicolumn{2}{l}{Hot-wire} & \multicolumn{2}{l}{Microphones} \\
\noalign{\smallskip}\cline{2-7}\noalign{\smallskip}
~& $Re_\tau$ & 2\,280 & Type & Dantec~55P15 & Type & GRAS\,46BE \\
~& $Re_\theta$ & 6\,190 & $l^+$ & 42.4 & $d^+$ & 13.6 \\
~& $\overline{U}_{\infty}$ & 14.8\,m/s & $l/d$ & 250 & $f_r^+$ & 0.15 \\
~& $\delta$ & 67.3\,mm & $\Delta T^+$ & 0.36 & $\Delta T^+$ & 0.36 \\
~& $u_{\tau}$ & 0.54\,m/s & $T \overline{U}_\infty/\delta$ & 32\,860 & $T \overline{U}_\infty/\delta$ & 32\,860 \\
~& $l^* = \nu/u_\tau$ & 29.5\,$\upmu$m & -- & -- & -- & -- \\
\end{tabular}
\caption{Parameters of data sets used: channel DNS data and experimental turbulent boundary layer data. Note that $\delta$ for DNS of the channel flows and the boundary layer experiment denote the channel halfwidth and the boundary layer thickness, respectively. {\scriptsize \dag Total simulation time without transition.}}
\label{tab:data} 
\end{center}
\end{table}

\begin{figure} 
\vspace{0pt}
\centering
\includegraphics[width = 0.999\textwidth]{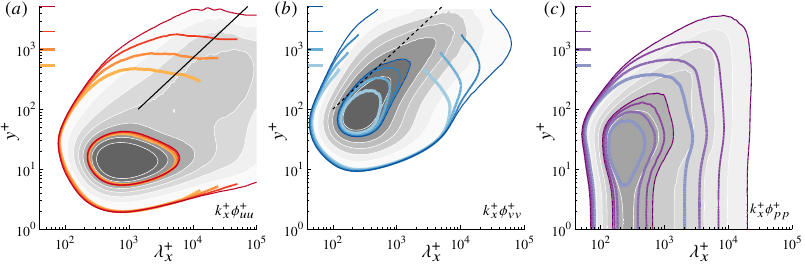}
\caption{1D spectrograms of ($a$) $u$, ($b$) $v$ and ($c$) $p$. Two clusters of solid, coloured iso-contours on (a-c) correspond to two contour values of $k_x^+\phi^+_{uu} = [0.2;~1.2]$, $k_x^+\phi^+_{vv} = [0.05;~0.3]$ and $k_x^+\phi^+_{pp} = [0.45;~2.25]$, respectively, for all $Re_\tau$ cases (an increased colour intensity corresponds to an increase in $Re_\tau$; the channel half-widths $\delta^+ = Re_\tau$ are indicated along the ordinate). The grey scale contour shows a finer discretization of iso-contours for the R5200 case only.}
\label{fig:spectro}
\end{figure}

For reference, 1D spectrograms of $u$, $v$ and $p$ are presented in figures~\ref{fig:spectro}($a$-$c$), with iso-contours of the pre-multiplied spectrum, \emph{e.g.}, $k_x^+\phi^+_{uu}$. Four sets of coloured iso-contours at the two values indicated in the caption correspond to the four Reynolds number cases, and the small lines at the ordinates correspond to each $Re_\tau \equiv \delta^+$. Only for the highest Reynolds number case, R5200, a finer discretization of grey-filled contours is shown. Data are presented in a similar manner later on in \S\,\ref{sec:yscaling}. 

As is well-known, the most energetic content resides at $y^+ \approx 15$ and $\lambda_x^+ \approx 10^3$ for the streamwise velocity, and $\lambda_x^+ \approx 250$ for the wall-normal velocity and wall-pressure. The pressure spectrum remains constant for $y^+ \lesssim 5$, reflecting the wall-pressure spectrum. For all fluctuating quantities in figure~\ref{fig:spectro}, their variance, at a given $y^+$, grows due to additional energy at large $\lambda_x^+$. For instance, see \citet{mathis:2009a} for the scaling of $u$-spectrograms and \citet{panton:2017a} for the scaling of the mean-square pressure fluctuations.

\subsection{Experimental wall-pressure--velocity data}\label{sec:dataexp}
\begin{figure} 
\vspace{0pt}
\centering
\includegraphics[width = 0.999\textwidth]{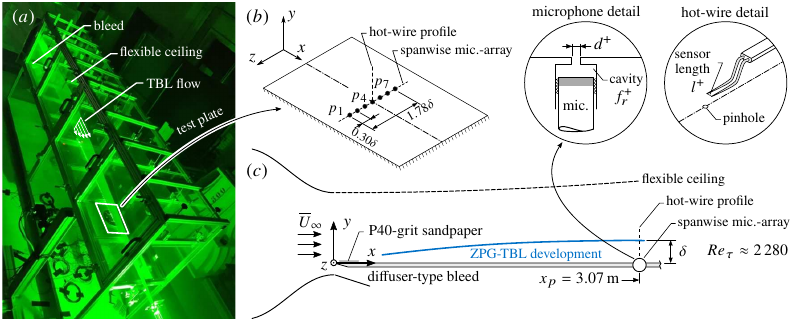}
\caption{($a$) Setup for the ZPG-TBL studies in the W-tunnel. ($b$,$c$) Experimental arrangement with a single hot-wire probe and a pinhole-mounted microphone array measuring the wall-pressure at 7 spanwise positions. In addition, one microphone is placed in the free-stream to measure the facility (acoustic) noise.}
\label{fig:exper}
\end{figure}

Experiments were carried out in the W-tunnel facility within the Faculty of Aerospace Engineering at the Delft University of Technology. This open-return tunnel has a contraction ratio of 4.7:1, with a square cross-sectional area of $0.60 \times 0.60$\,m$^2$ at the inlet of the test section, and can produce a maximum flow velocity of roughly $16.5$\,m/s. For generating a TBL flow, a setup was developed with a relatively long streamwise development length, consisting of a metal frame with poly-carbonate walls for optical access (figure~\ref{fig:exper}$a$). The bottom wall has a width of 0.60\,m and a total length of 3.75\,m. The TBL was initiated just downstream of the leading edge at $x = 0$ with a trip of P40-grit sandpaper over a length of 0.115\,m and over the full perimeter. The floor was suspended 0.08\,m above the bottom wall of the upstream tunnel contraction, allowing for a clean start of the TBL flow without leading-edge separation. A flexible, 4\,mm thick poly-carbonate ceiling was configured for a zero-pressure gradient (ZPG) development of the TBL. This ZPG condition was assessed using the acceleration parameter $K = (\nu/U_e^2)(\mathrm{d}\overline{U}_e/\mathrm{d}x)$, where $\overline{U}_e(x)$ is the boundary layer edge-velocity. A Pitot-static tube in the potential flow at $x_p = 3.07$\,m provided measures of the total ($p_0$) and static ($p$) pressures, and thus $\overline{U}_\infty = \overline{U}_e$ at $x_p$ (the static temperature and barometric pressure were also recorded for inferring the air density and viscosity). Using two streamwise rows of (in total) 100 static pressure taps in the floor, providing $p(x)$, the variation in $\overline{U}_e(x)$ was computed as $\overline{U}_e(x) = \sqrt{2\left(p_0 - p(x)\right)/\rho}$, given that the total pressure remains constant along a streamline. For the nominal free-stream velocity of the current study ($\overline{U}_\infty \approx 15$\,m/s), the acceleration parameter remained within an acceptable range for a ZPG condition along the entire test section \citep[$K < 1.6 \cdot 10^{-7}$ following][]{schultz:2007a}. Finally, measurements are performed near the aft of the setup around $x_p = 3.07$\,m. Here, the free-stream turbulence intensity at the nominal free-stream velocity, based on the hot-wire measurement described later, is $\sqrt{\overline{u^2}}/\overline{U}_\infty \approx 0.3$\%.

Hot-wire anemometry measurements were performed with a Dantec Dynamics 55P15 miniature-wire boundary layer probe. This single-wire probe only yields the streamwise velocity (given that $u \gg v$ in boundary layer flows) and comprised a plated Tungsten wire with a diameter of $d_w = 5\,\upmu$m and a sensing length of $l = 1.25$\,mm (resulting in $l/d_w = 250$). The viscous-scaled wire-length of $l^+ = 42.4$ yields an acceptable spatial resolution \citep{hutchins:2009a}, given that this study concentrates primarily on the logarithmic region for wall-pressure--velocity correlations (the measurement is fully-resolved in that region, as shown later on). Hot-wire traversing was done with a Zaber X-LRQ300HL-DE51 traverse, with an integrated encoder and controller yielding a positional accuracy of $13\,\upmu$m (a resolution better than $0.4l^*$). A Taylor-Hobson micro alignment telescope was used to position the hot-wire at the most near-wall position before performing a wall-normal traverse spanning 40 logarithmically spaced positions in the range $10.2 \lesssim y^+ \lesssim 1.2Re_\tau$.

The hot-wire was operated in constant temperature mode using a TSI IFA-300 bridge at an overheat ratio of 1.8. For each hot-wire position, signals were sampled (simultaneously with the wall-pressure sensors) at a rate of $\Delta T^+ \equiv u_\tau^2/\nu/f_s = 0.36$, where $f_s = 51.2$\,kHz is the sampling frequency. The anemometer system low-pass filtered the voltage signal with a spectral cut-off at $20$\,kHz. Sampling was done with a 24-bit A/D conversion through an NI\,9234 module embedded in an NI\,cDAQ. Relatively long signals were acquired with an uninterrupted acquisition time of $T_a = 150$\,seconds at each wall-normal position, resulting in more than $32\,860$ boundary layer turnover times; this is more than sufficient for converged spectral statistics at the lowest frequencies of interest. An \emph{in-situ} calibration of the hot-wire was performed in the potential flow-region using the reference velocity provided by the Pitot-static tube. A correction method for hot-wire voltage drift due to variations in barometric pressure and static temperature was also implemented \citep{hultmark:2010a}.

\begin{figure} 
\vspace{0pt}
\centering
\includegraphics[width = 0.999\textwidth]{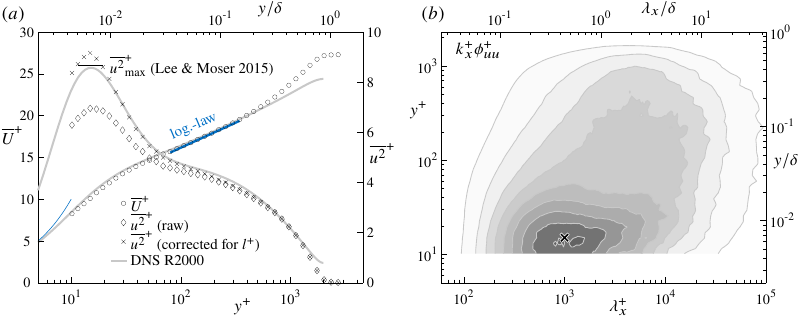}
\caption{($a$) Experimental boundary layer profiles of the streamwise mean velocity and turbulence intensity, compared to the DNS R2000 case. The mean velocity profile is compared to the log.-law with constants $\kappa = 0.384$ and $B = 4.17$, and the $u$-TKE profile is corrected for spatial resolution effects. ($b$) Premultiplied energy spectrogram $k^+_x\phi^+_{uu}$ (filled iso-contours 0.2:0.2:1.8); the scale-axis is converted to a wavelength-dependence using $\lambda_x \equiv \overline{U}(y)/f$.}
\label{fig:TBLprofs}
\end{figure}

Profiles of $\overline{U}$ and $\overline{u^2}$ are plotted in figure~\ref{fig:TBLprofs}($a$) and the boundary layer parameters are listed in table~\ref{tab:data}. These parameters were obtained by fitting the mean velocity profile to a composite profile with log-law constants of $\kappa = 0.384$ and $B = 4.17$ \citep{chauhan:2009a}. On the basis of these values, $Re_\tau = 2\,280$ and the Reynolds number based on the momentum thickness is $Re_\theta \equiv \overline{U}_\infty \theta/\nu = 6\,190$. A local friction coefficient of $C_f = 2u_\tau^2/\overline{U}_\infty^2 \approx 2.69\cdot 10^{-3}$ matches a Coles-Fernholtz relation, $C_f = 2\left[\nicefrac{1}{0.38}\ln\left(Re_\theta\right) + 3.7\right]^{-2}$, to within 4.5\%. The mean velocity profile compares well to the R2000 case of the DNS, up to the wake-region. From the streamwise turbulence intensity profile, it is clear that the experimental data are attenuated due to the hot-wire's spatial resolution. When correcting for this limited resolution via the method of \citet{smits:2011HWa}, it matches well with the DNS profile in the buffer region and above, but a slight overestimate of the expected peak-value at $y^+ \approx 15$ is observed, following $\overline{u^2}_{\rm max} = 0.63\ln\left(Re_\tau\right) + 3.80$ \citep{lee:2015a,smits:2021a}. We ascribe this mismatch to wall-proximity effects \citep{hutchins:2009a}.

Prior to computing frequency spectra $\phi_{uu}(f)$, the experimental time series $u(y,t)$ were down-sampled with a factor of 5 to match the corrected wall-pressure data (see Appendix~\ref{sec:appA}). One-sided spectra $\phi_{uu}(y;f) = 2\langle U(y;f) U^*(y;f)\rangle$ were computed by way of ensemble-averaging a total of 180 FFT partitions of $N = 2^{14}$ samples (50\,\% overlap and a Hanning window applied to them). This yields a spectral resolution of ${\rm d}f = 0.625$\,Hz or ${\rm d}f^+ = 3.4 \cdot 10^{-5}$. For interpretive purposes, frequency spectra are converted to wavenumber spectra using a single convection velocity $\overline{U}_c$ (taken as the local mean velocity $\overline{U}(y)$ unless stated otherwise). With wavenumber $k_x = 2\pi f/\overline{U}_c$ the wavenumber spectrum becomes $\phi_{uu}(k_x) = \phi_{uu}(f){\rm d}f/{\rm d}k$, where group ${\rm d}f/{\rm d}k = \overline{U}_c/(2\pi)$ converts the energy density from a `per unit frequency' to a `per unit wavenumber' (the spectral resolution becomes ${\rm d}k \approx 0.88$\,m$^{-1}$ or ${\rm d}k^+ \approx 2.6 \cdot 10^{-5}$ at the lowest position of $y^+ = 10$). Throughout this paper, the scale axis is either presented in terms of $k_x$ or wavelength $\lambda_x = 2\pi/k_x$. Figure~\ref{fig:TBLprofs}($b$) presents the streamwise energy spectrogram $k_x\phi_{uu}$ for validation of the experimental setup. The inner-spectral peak is clearly visible and is identified with the $\times$ marker at $\lambda_x^+ = 10^3$ and $y^+ = 15$. Larger scales are more energetic in the logarithmic region, although the Reynolds number is not yet high enough for a discernible outer-spectral peak to appear \citep{baars:2020P1a}.

Wall-pressure measurements were simultaneously made with the hot-wire ones, and thus sampled with the parameters listed in table~\ref{tab:data}. Eight GRAS\,46BE $\nicefrac{1}{4}$\,in. CCP free-field microphones were employed. Seven of them formed a spanwise array for wall-pressure measurements, while one was used in the potential flow region to measure facility noise. The microphone sets have a nominal sensitivity of $3.6$\,mV/Pa and a frequency response range with an accuracy of $\pm 2$\,dB for $4$\,Hz to $80$\,kHz, while for the range $10$\,Hz to $40$\,kHz the accuracy is $\pm 1$\,dB. The dynamic range is $35$\,dB to $160$\,dB (with a reference pressure of $p_{\rm ref} = 20\,\upmu$Pa). For our current wall-pressure--velocity correlation study, the primary frequencies of interest lay between roughly $5$\,Hz and $800$\,Hz, and the measured pressure intensity is on the order of $105$\,dB, thus making these microphones suitable for these types of measurements.

The spanwise array of seven equally-spaced pinhole-mounted microphones had an inter-spacing of $20$\,mm, or $0.30\delta$, with the total width spanning $\Delta z = 1.78\delta$. The hot-wire profile was measured above the center pinhole. Each microphone was screwed inside a cavity (after removal of the microphone grid-cap) so that the sensing diaphragm formed the bottom of the cavity. On the back side, underneath the wind tunnel floor, a box surrounding the microphones prevented any pressure fluctuations at their venting holes. On the TBL side, a pinhole with a diameter of $d^+ = 13.6$ ($d = 0.40$\,mm) ensured a sufficient spatial resolution of the measurement \citep{gravante:1998a}. The pinhole depth was $t = 0.80$\,mm and the cavity diameter matched the microphone-body outer diameter ($D = 6.35$\,mm). The cavity length was designed as $L = 2.0$\,mm, so that the Helmholtz resonance frequency of the cavity was above the frequency range of interest ($f_r^+ = f_r\nu/u_\tau^2 = 0.15$ or $f_r = 2\,750$\,Hz). Raw signals of the pinhole-microphone measurements required post-processing to yield valid time series of the wall-pressure fluctuations. The post-processing steps are described in Appendix~\ref{sec:appA}.

\section{Scaling of the wall-pressure--velocity coupling}\label{sec:yscaling}
This section utilizes the DNS data to assess the coupling between the fluctuations of $u$ and $v$, and the wall-pressure field $p_w$. First, we proceed with a 1D spectral analysis in the streamwise direction, which is reminiscent of the data available from typical experiments.

\subsection{1D analysis in the streamwise direction}\label{sec:1Dscaling}
\begin{figure} 
\vspace{0pt}
\centering
\includegraphics[width = 0.999\textwidth]{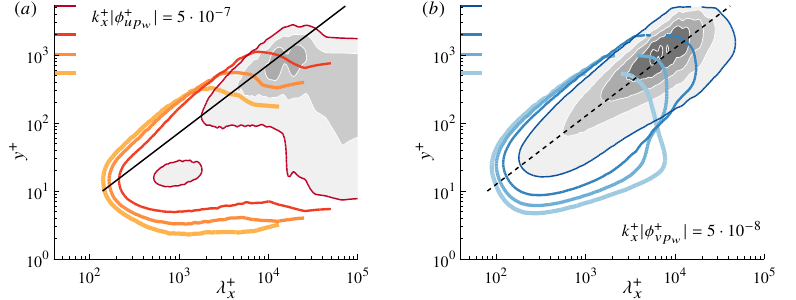}
\caption{($a$) 1D gain of the cross-spectrogram, computed using $u$ and $p_w$. Solid, coloured iso-contours correspond to one contour value indicated in the figure for all $Re_\tau$ cases (an increased colour intensity corresponds to an increase in $Re_\tau$); the greyscale contour shows a finer discretization of iso-contours for the R5200 case only. A black trend line indicates a wall-scaling of $\lambda_x/y = 14$. ($b$) Similar to sub-figure ($a$), but now for $v$ and $p_w$. A black-dashed trend line indicates a wall-scaling of $\lambda_x/y = 8$.}
\label{fig:1Dcross}
\end{figure}

Cross-spectra of wall-pressure and velocity yields an indication of the coupling of absolute energy. We only examine the gain of the complex-valued 1D cross-spectrum, $\phi_{up_w}\left(\lambda_x,y\right)$; the phase is beyond the scope of our current work and is only relevant for spatial/temporal lags. The gain of the cross-spectrogram is presented with iso-contours of $\vert\phi^+_{up_w}\vert$ in figure~\ref{fig:1Dcross}($a$). One particular contour value is chosen and the iso-contours correspond to the four Reynolds number cases of the DNS (table~\ref{tab:data}). It is evident that the region of cross-spectral energy grows along the black solid line with an increase in $Re_\tau$. Since the black line indicates a constant ratio of $\lambda_x/y$, this trend is representative of distance-from-the-wall scaling.

\begin{figure} 
\vspace{0pt}
\centering
\includegraphics[width = 0.999\textwidth]{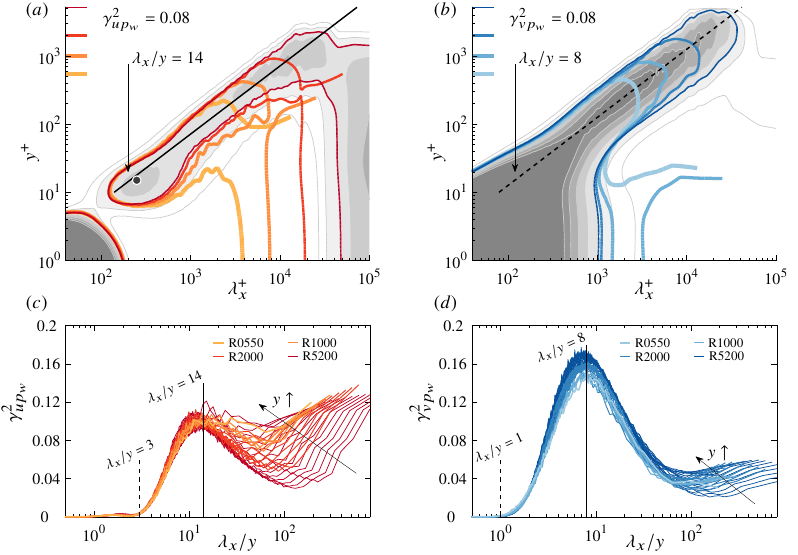}
\caption{($a$) Linear coherence of $u$ and $p_w$. Solid, coloured iso-contours correspond to one contour value for all $Re_\tau$ cases (an increased colour intensity corresponds to an increase in $Re_\tau$); the greyscale contour shows a finer discretization of contours for $Re_\tau \approx 5\,200$ only. ($b$) Similar to sub-figure ($a$), but now for the linear coherence of $v$ and $p_w$. ($c$) Linear coherence spectra $\gamma^2_{up_w}$ (one bundle of lines per $Re_\tau$ condition) within the logarithmic region, in the range $80 \lesssim y^+ \lesssim 0.15Re_\tau$. Results in frequency space are converted to a wavelength-dependence using $\lambda_x \equiv \overline{U}(y)/f$. ($d$) Similar to sub-figure ($c$), but now with the linear coherence spectra $\gamma^2_{vp_w}$.}
\label{fig:1Dcohlin}
\end{figure}

\begin{figure} 
\vspace{0pt}
\centering
\includegraphics[width = 0.999\textwidth]{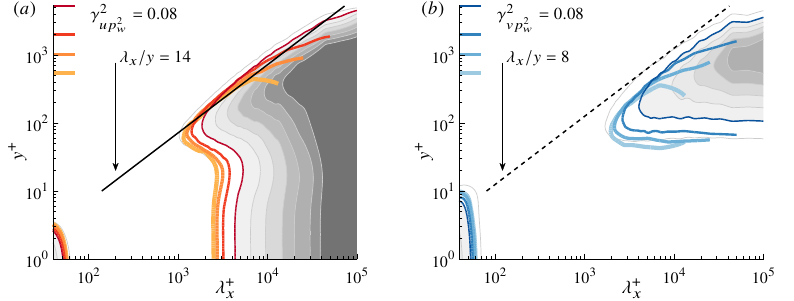}
\caption{Similar to \ref{fig:1Dcohlin}, but now for ($a$) $u$ and $p^2_w$ (wall-pressure squared), and ($b$) $v$ and $p^2_w$.}
\label{fig:1Dcohnon}
\end{figure}

A normalized coherence is now considered to explore how the coupling scales, independent of the scaling of $u$ energy). The coherence
\begin{equation}\label{eq:lcs1d}
    \gamma^2_{up_w}\left(\lambda_x,y\right) = \frac{\vert \phi_{up_w}\left(\lambda_x,y\right)\vert^2}{\phi_{uu}\left(\lambda_x,y\right)\phi_{p_wp_w}\left(\lambda_x\right)},
\end{equation}
is presented in figure~\ref{fig:1Dcohlin}($a$) using one iso-contour of $\gamma^2_{up_w}$ for all four $Re_\tau$ of the DNS. For the highest Reynolds number case, R5200, grey-filled contours show that a spectral band of strong coherence appears at a self-similar scaling of $\lambda_x/y \approx 14$; this scaling also appears to be Reynolds-number invariant. To further accentuate these observations, individual coherence spectra are superimposed in figure~\ref{fig:1Dcohlin}($c$). Here, each bundle of lines with the same colour corresponds to one Reynolds-number dataset. Coherence spectra are plotted corresponding to a relatively coarse grid of $y$ positions that are logarithmically spaced between lower- and upper bounds of an extended logarithmic region, chosen as $y^+ = 80$ and $y^+ = 0.16Re_\tau$, respectively (resulting in 1 curve for the R0550 data, and 15 curves for the R5200 data). A wall-scaling of $\lambda_x/y$ is adopted on the abscissa. It is evident that the coherence spectra collapse; signifying a Reynolds-number invariant location of the coherence peak and a constant coherence magnitude. This coherence-peak location at $\lambda_x/y  = 14$ closely matches the Reynolds number invariant aspect ratio of wall-attached coherent structures of $u$, see \citet{baars:2017a}. The relatively low magnitude indicates a weak linear coherence, but this is expected given that (extreme) events in the wall-pressure are primarily caused by sweeps, ejections, and thus motions associated with strong vertical velocity fluctuations (and less so with $u$ fluctuations). This can, for instance, be observed in the work by \citet{ghaemi:2013a} where high amplitude near-wall pressure events are considered by visualizing the associated conditional pressure and velocity structures. In this regard, it is evident from figure~\ref{fig:1Dcohnon}($a$) that the `self-similar scaling' of coherence appears at its smallest scale near the wall with a peak at $y^+ \approx 15$ and $\lambda_x^+ \approx 14 \times 15 \approx 210$ (visualized with a round marker). This streamwise scale agrees with the typical (conditional) near-wall turbulent structure associated with these high-amplitude pressure peaks observed by \citet{ghaemi:2013a}.

By further inspection, it is intriguing to note the appearance of a region of zero-coherence at relatively large wavelengths and at positions below the ridge, even though this region resides closer to the wall. It can thus be concluded that large-scale $u$ fluctuations in close proximity to the wall do not comprise a phase-consistent linear relationship with the wall-pressure, whereas the $u$ fluctuations further up in the wall-bounded turbulent flow do. Presumably, this is due to the stronger dispersive (and random) nature of the convection of large-scale structures near the wall \citep{liu:2020a}. The wall-pressure footprint for each scale is thus only linearly coupled to $u$ fluctuations that are statistically self-similar and that govern the footprint from each respective height. This is consistent with the fact that the pressure scalar is influenced by the velocity field in the entire domain; this is opposite to, for instance, wall-attached velocity fluctuations that are coherent throughout the full wall-normal extent \citep{baars:2017a}.

Finally, two dominant regions of non-zero coherence appear: (1) very near the wall at wavelengths $\lambda_x^+ \lesssim 200$ and at $y^+ \lesssim 5$, and (2) at large outer-scaled wavelengths (\emph{e.g.}, for $Re_\tau \approx 5\,200$ for wavelengths $\lambda_x^+ \gtrsim 3 \cdot 10^4$). Although these regions have a non-zero coherence, it is important to note that $u$ has no significant energy in region 1 (figure~\ref{fig:spectro}$a$) and neither has $p_w$ in region 2 (figure~\ref{fig:spectro}$c$). Likewise, the cross-spectral energy is low in these regions (recall figure~\ref{fig:1Dcross}$a$). Although insignificant in terms of absolute energy, very large \emph{global modes} of velocity fluctuations spanning the entire wall-normal extent (and simulation box) can be responsible for the large-scale coherence in region 2 \citep{delalamo:2003a,jimenez:2008a}.

A correlation analysis for $p_w$ with $v$ (instead of $u$) results in the spectrograms of $\vert \phi_{vp_w}\vert$ and $\gamma^2_{vp_w}$ shown in figures~\ref{fig:1Dcross}($b$) and~\ref{fig:1Dcohlin}($b$), respectively. As for $p_w$ and $u$, coherence spectra for $p_w$ with $u$ are superimposed in figure~\ref{fig:1Dcohlin}($d$), as function of $\lambda_x/y$ and for a range of $y$ positions for all Reynolds-number datasets, to bring attention to the self-similar and Reynolds number-invariant coherence-peak near $\lambda_x/y = 8$. As for the $u$ fluctuations, a significant bulk of the coherence resides at small wavelengths very near the wall when $p_w$ and $v$ are concerned, but this coherence is insignificant given the absence of cross-spectral energy (figure~\ref{fig:1Dcross}$b$). The main finding that the wall-pressure-coherent energy in $v$ resides at smaller wavelengths than their streamwise counterpart is interpreted as follows. Fluctuations in $v$ induce wall-pressure fluctuations through a `flow stagnation' when directed towards the wall. When wall-pressure coherent velocity fluctuations are thought of as wall-attached motions (\emph{e.g.}, hairpins or packets of them), their centre regions contain $u < 0$, while the $v$-fluctuations induced by the vortical motions of vortex heads inducing wall-pressure reside at shorter $\lambda_x$ (statistically thus nearly half the size). This is evident from spectra: while the peak in the $uv$ cospectra resides around $\lambda_x/y \approx 15$ (dominated by the higher energy in $u$, residing at relatively large streamwise scales), the dominant energy in the spectra of $v$ alone appears at much shorter scales throughout the logarithmic region, along $\lambda_x/y \approx 2$ \citep{baidya:2017a}. In addition, \citet{jimenez:2008a} showed that pressure spectra scale relatively well with the local Reynolds shear stress $\overline{uv}^2$ (representative of the intensity of eddying motions). Given the relatively large streamwise scales in $\overline{uv}$ and the fact that the wall-pressure spectrum embodies a footprint of the global pressure fluctuations, the peak-coherence of $p_w$ with $v$ does indeed reside at slightly larger scales $\lambda_x/y \approx 8$ \citep[than the peak location of the $v$ spectra at $\lambda_x/y \approx 2$, see][]{baidya:2017a}.

Higher order terms of the wall-pressure--velocity coupling are of relevance to the analysis in \S\,\ref{sec:state}, when $p_w$ forms the input for estimates of the velocity fluctuations. For this reason, coherence spectra of the wall-pressure squared, with both the $u$ and $v$ velocities, are presented as spectrograms in figures~\ref{fig:1Dcohnon}($a$) and~\ref{fig:1Dcohnon}($b$), respectively. It is important to note here that $p_w^2$ is the square of the de-meaned wall-pressure (this is \emph{not} the same as the de-meaned wall-pressure-squared). Note that this form of a nonlinear correlation does address interactions of different scales (\emph{e.g.} an interscale interaction of pressure with velocity), although not explicitly in terms of triadic (or higher-order) scale interactions. Bipectral analysis is required to address those interactions \citep{baars:2014a,cui:2021a}, which is beyond the scope of this manuscript. Focusing on the coherence of $p_w^2$ with $u$, \citet{naguib:2001a} had hypothesized that this quadratic pressure-interaction represents a flow structure obeying outer-scaling. Trends of $\gamma^2_{up_w^2}$ show a self-similar scaling in the logarithmic region ($y^+ \gtrsim 100$) and coherence only appears for $\lambda_x \gtrsim 14y$; note that this corresponds to the region where $\gamma^2_{up_w}$ starts to decrease from its peak-ridge at $\lambda_x/y = 14$. A distance-from-the-wall scaling is also seen in the trends of $\gamma^2_{vp_w^2}$, which appears to become relevant at scales slightly larger than $\lambda_x/y = 8$ (the lowest locations at which $p_wv$ coherence appears is not fixed in outer-scaling). 

Physically, the coherence at relatively large scales involving $p_w^2$ would be reminiscent of nonlinearities associated with an intensity-modulation phenomenon \citep{tsuji:2007a}. This modulation of the near-wall quantities in wall-bounded turbulence is induced by large-scale velocity fluctuations that are most energetic in the logarithmic region (and are most pronounced at high values of $Re_\tau$) and that leave a direct imprint on the wall. This imprint changes the local, large-scale friction velocity and, thus the viscous scale. Consecutively, this yields modulated near-wall pressure and small-scale velocity fluctuations given that these near-wall quantities are universal in viscous scaling \citep{zhang:2016a,chernyshenko:2021a}. Thus, when the intensity of (wall-)pressure fluctuations are modulated by the large-scale $u$ (or $v$) fluctuations, the intensity-modulation `envelope' (which by itself has no energy contribution in $p_w$) becomes energetic in the square of the de-meaned wall-pressure signal, $p_w^2$. At the same time, the usage of $p_w^2$ results in linear coherence at wavelengths larger than where the coherence peaks with the linear pressure term (figure~\ref{fig:1Dcohnon}$a$,$b$). Further inspection of figure~\ref{fig:1Dcohnon} reveals that the coherence is strongest for the velocity fluctuations taken in the logarithmic region; this reflects the general consensus in the community that the modulation of wall-quantities is driven by energetic large-scale motions in the logarithmic region of the flow. As a final note, we confirmed that the coherence with any higher-order pressure terms ($p_w^3$ with either $u$ or $v$) was zero.

\subsection{2D analysis in the streamwise--spanwise plane}\label{sec:2Dscaling}
\begin{figure} 
\vspace{0pt}
\centering
\includegraphics[width = 0.999\textwidth]{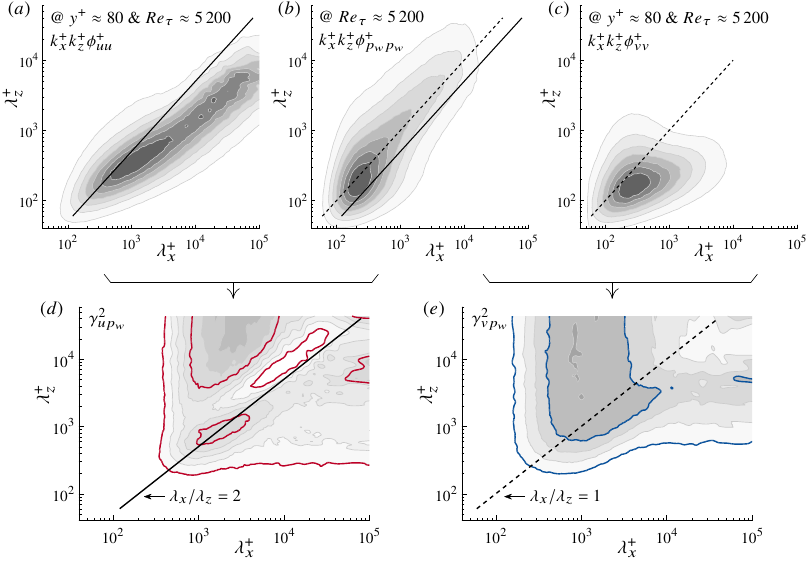}
\caption{2D spectrograms of ($a$) $u$, ($b$) $p_w$ and ($c$) $v$ at one $Re_\tau$ and $y^+$, as indicated in the sub-figures; grey-filled contours show 7 iso-contours ranging up to the maximum value. ($d$) 2D linear coherence of $u$ and $p_w$, for the same $Re_\tau$ and $y^+$ as ($a$,$b$); the solid, coloured iso-contours correspond to two contour values of $\gamma^2_{up_w} = [0.05;~0.25]$. ($e$) Similar to sub-figure ($d$), but now for the 2D linear coherence of $v$ and $p_w$; the solid, coloured iso-contours correspond to two contour values of $\gamma^2_{vp_w} = [0.10;~0.30]$.}
\label{fig:2Dcoh-nor}
\end{figure}

In the 1D spectral analysis, all spanwise information was lumped together. To inspect the full spectral picture in both homogeneous directions, we move towards a 2D analysis. At first the 2D spectrograms of the streamwise velocity, $\phi_{uu}\left(\lambda_x,\lambda_z,y\right)$, and wall-pressure, $\phi_{p_wp_w}\left(\lambda_x,\lambda_z\right)$, are shown for $y^+ \approx 80$ at $Re_\tau \approx 5\,200$ in figures~\ref{fig:2Dcoh-nor}($a$) and~\ref{fig:2Dcoh-nor}($b$), respectively. A solid black line in figure~\ref{fig:2Dcoh-nor}($a$) indicates a scaling of $\lambda_x/\lambda_z = 2$ whereas the dashed line in figure~\ref{fig:2Dcoh-nor}($b$) signifies a scaling of $\lambda_x/\lambda_z = 1$. Energy in $u$ at larger wavelengths is more anisotropic at this position in the near-wall region and constitutes larger streamwise scales than spanwise scales. This is reminiscent of (very) large-scale motions in the streamwise direction. The wall-pressure reflects a more isotropic behaviour in the two homogeneous directions, owing to the fact that wall-pressure has a stronger coupling with the $v$ motions, which are themselves more isotropic than $u$ motions. Nevertheless, the scope of the current paper is to inspect the \emph{normalized} coherence, since this reveals the degree of phase-consistency and thus the energy-fraction of the velocity fluctuations that are stochastically coupled to the wall-pressure. Coherence in 2D is a generalization of \eqref{eq:lcs1d}, following:
\begin{equation}\label{eq:lcs2d}
    \gamma^2_{up_w}\left(\lambda_x,\lambda_z,y\right) = \frac{\vert \phi_{up_w}\left(\lambda_x,\lambda_z,y\right)\vert^2}{\phi_{uu}\left(\lambda_x,\lambda_z,y\right)\phi_{p_wp_w}\left(\lambda_x,\lambda_z\right)} \in [0,1].
\end{equation}
Coherence for the data used to present the 2D (auto-)spectrograms in figures~\ref{fig:2Dcoh-nor}($a$,$b$) is shown in figure~\ref{fig:2Dcoh-nor}($d$). Notably, the coherence only appears for $\lambda_x^+ \gtrsim 400$ (when taking a threshold of $\gamma^2_{up_w} = 0.05$), agreeing to what was observed from the 1D analysis in figure~\ref{fig:1Dcohlin}. A coherence-peak in the 2D view still resides close to $\lambda_x/y \approx 14$ (for figure~\ref{fig:2Dcoh-nor}($d$) equivalent to $\lambda_x^+ = 14y^+ = 1120$) and is reasonably symmetric around $\lambda_x/\lambda_z = 2$. The latter implies that wall-pressure--coherent $u$ motions are roughly twice as long as they are wide. Another region of peak-coherence resides at very large $\lambda_z$, but the energy there is insignificant and hence this high coherence is physically irrelevant. For instance, at $(\lambda_x^+,\lambda_z^+) = (10^3,10^4)$ in figure~\ref{fig:2Dcoh-nor}($d$), $\gamma^2_{up_w} \approx 0.25$ even though the absolute energy of $u$ is invisible in figure~\ref{fig:2Dcoh-nor}($a$). For the $v$ motions (auto-spectrogram shown in figure~\ref{fig:2Dcoh-nor}$c$) the 2D coherence in figure~\ref{fig:2Dcoh-nor}($e$) shows symmetry around $\lambda_x/\lambda_z = 1$. A further statistical interpretation is reserved for when the 2D coherence is presented in a different manner, described next.

\begin{figure} 
\vspace{0pt}
\centering
\includegraphics[width = 0.999\textwidth]{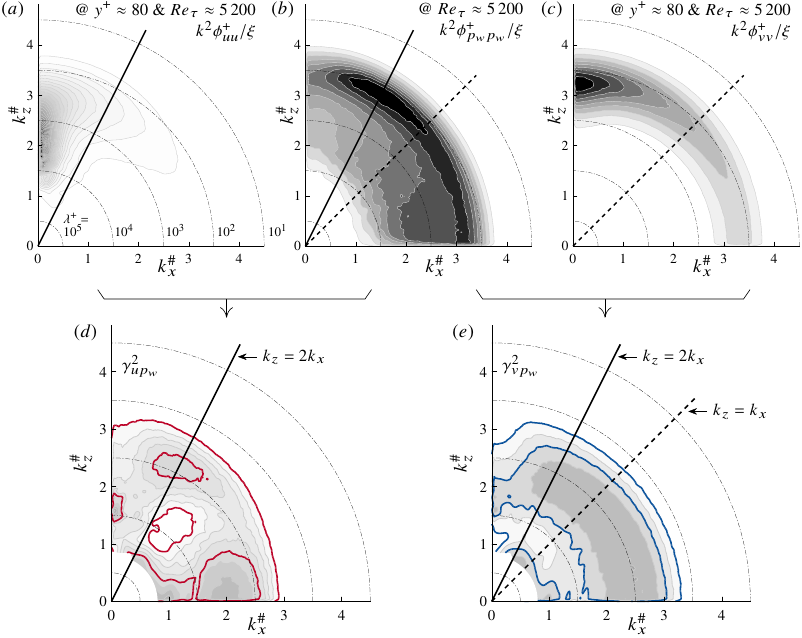}
\caption{($a$,$b$,$c$) Similar to figures~\ref{fig:2Dcoh-nor}($a$,$b$,$c$) but now presented in log-polar format as described in the text; grey scale contours show 7 iso-contours ranging up to the maximum value. ($d$,$e$) Similar to figures~\ref{fig:2Dcoh-nor}($d$,$e$) but now presented in log-polar format. For indicating trends of scale-isotropy, five dash-dotted arcs of constant $k = 2\pi/\lambda$ are plotted in all sub-figures and correspond to $\lambda^+ = 10^1$, $10^2$, $10^3$, $10^4$ and $10^5$ (from the most outer-arc going inward).}
\label{fig:2Dcoh-log}
\end{figure}

So far, 2D auto-spectrograms were pre-multiplied with $k_xk_z$ and visualized with iso-contours of $k_xk_z\phi$ in a log-Cartesian space. This visualization ensures that the `area-under-the-contour' is proportional to the energy residing in logarithmic ranges of $\lambda_x$ and $\lambda_z$. \citet{lee:2019a} pointed out that this method of visualization also comes with two primary shortcomings. In summary, the distortion of the 2D wavevector orientation complicates observing the alignment of modes with the $x$ and $z$ directions: all lines of constant $\lambda_x/\lambda_z$ (or $k_z/k_x$) have a slope of 1 and are only marginally offset on the log-log plot, for different constants of $\lambda_x/\lambda_z$ (see figure~\ref{fig:2Dcoh-nor}$b$). In addition, vector wavelengths with a similar magnitude $\lambda = \left(\lambda_x^2 + \lambda_z^2\right)^{\nicefrac{1}{2}}$ follow a non-trivial trend line that creates difficulties in assessing scale isotropy. That is, when turbulence quantities contain an equal energy content at vector wavelengths (or wavenumbers) that have the same \emph{magnitude} in the wall-parallel plane, this energy is said to be distributed isotropically in scale. Because of the two shortcomings, \citet{lee:2019a} introduced a `log-polar' format for the 2D spectrum. When considering wavenumbers, $\phi\left(k_x,k_z\right)$ can be expressed in polar coordinates, yielding $\phi\left(k\cos\theta,k\sin\theta\right)$ with $k_x = k\cos\left(\theta\right)$ and $k_z = k\sin\left(\theta\right)$. When defining $\xi = \log_{10}\left(k/k_{\rm ref}\right)$, where $k_{\rm ref}$ is a reference wavenumber smaller than the smallest non-zero wavenumber considered, a set of logarithmic polar coordinates would be $\xi$ and $\theta$, with associated Cartesian coordinates $k_x^\# = \xi \cos\theta = \xi k_x/k$ and $k_z^\# = \xi \sin\theta = \xi k_z/k$, respectively. Following Parseval's theorem, the spectral-integral equals
\begin{equation}\label{eq:logp}
    \int_{k_x>0}\int_{k_z>0} \phi\,{\rm d}k_x{\rm d}k_z = \int_{\theta=0}^{\pi/2}\int_{\xi>0} \frac{k^2}{\xi}\phi\,\xi\,{\rm d}\xi{\rm d}\theta = \int_{k^\#_x>0}\int_{k^\#_z>0} \frac{k^2}{\xi}\phi\,{\rm d}k^\#_x{\rm d}k^\#_z.
\end{equation}
Hence, to preserve the total energy in plotting, quantity $k^2\phi/\xi$ is visualized with linear axes $k_x^\#$ and $k_z^\#$. Hence, lines of constant $k_z/k_x$ possess slopes of $k_z/k_x$, and a curve of constant $k$ becomes an arc. Figure~\ref{fig:2Dcoh-log} re-plots all 2D spectrograms of figure~\ref{fig:2Dcoh-nor} using the log-polar format, with $k^+_{\rm ref} = 1/50\,000$. Five arcs of constant $k = 2\pi/\lambda$ indicating the trend of scale-isotropy are visualized with the dash-dotted lines on all plots and correspond to $\lambda^+ = 10^1$, $10^2$, $10^3$, $10^4$ and $10^5$ from the largest to the smallest arc. Lines of constant $k_z/k_x$ correspond to the $k_z = 2k_x$ trend (solid line) and $k_z = k_x$ trend (dashed line).

For a detailed interpretation of the 2D energy spectrograms of $u$ and $v$ in log-polar format (figures~\ref{fig:2Dcoh-log}$a$,$c$), we refer to \citet{lee:2019a}. From the wall-pressure spectrogram (figure~\ref{fig:2Dcoh-log}$b$) it becomes evident that a significant part of the energy is distributed close to isotropically in scale, with a ridge at $\lambda^+ \approx 200$. The peak energy resides close to $\lambda_x = 2\lambda_z$, meaning that the pressure modes are, statistically, slightly elongated in the streamwise direction. Larger-scale energy (\emph{e.g.}, around the arc of $\lambda^+ = 10^3$) tends to be concentrated towards the $k_x^\#$ axis and thus resides in Fourier modes with $k_z/k_x$ lower than 1 (or $\lambda_x/\lambda_z$ lower than 1). When bringing in the coherence, for the wall-pressure with $v$ in figure~\ref{fig:2Dcoh-log}($e$), it peaks with a trend of scale-isotropy around the $\lambda^+ = 10^3$ arc for $p_wv$ structures that are not strongly elongated in the streamwise direction (residing at $k_z < 2k_x$); contrary, the energy in $v$ is relatively low at these large scales (figure~\ref{fig:2Dcoh-log}$c$). Because the coherence is relatively strong below the $k_z = k_x$ line (in fact, constant at vector wavenumbers with the same magnitude), while it weakens beyond $k_z \gtrsim 2k_x$, the following can be concluded. Fluctuations in $v$ that are elongated in the spanwise direction $z$ (residing at $k_z \lesssim k_x$) are energetically weak (see figure~\ref{fig:2Dcoh-log}$c$) but are coupled to the wall-pressure imprint more strongly than $v$ motions that are strongly elongated in $x$ (residing at $k_z \gtrsim 2k_x$). The former would represent coherent spanwise vorticity, while the latter is representative of coherent streamwise vorticity. Concerning the $u$ motions, $\gamma^2_{up_w}$ and the wall-pressure energy peak for streamwise elongated scales, along $k_z = 2k_x$. The scale at which the coherence peaks is larger since only the larger scales are coherent. Note that the single wall-normal position considered here is near $y^+ = 80$ and does not provide a full-view: lower positions cause the peak-coherence to reside at smaller scales (as was already observed in the 1D analysis, see figure~\ref{fig:1Dcohlin}$a$). Hence it will be beneficial to consider the scaling of coherence with $y$ and $Re_\tau$, before drawing further conclusions.
\begin{figure} 
\vspace{0pt}
\centering
\includegraphics[width = 0.999\textwidth]{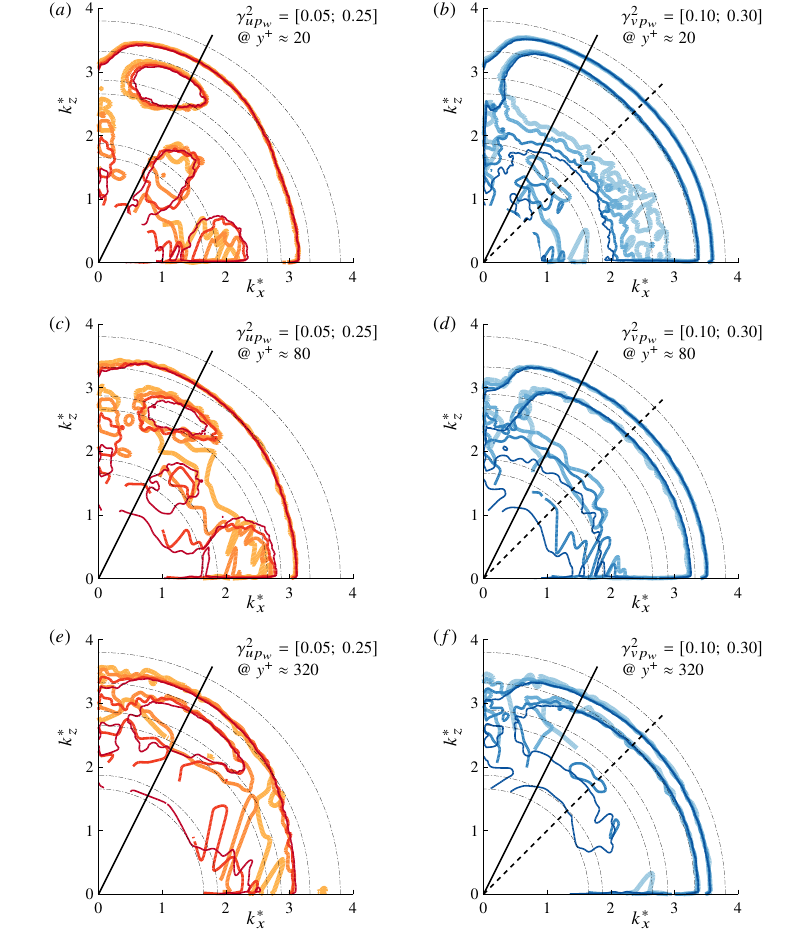}
\caption{($a$,$c$,$e$) 2D linear coherence of $u$ and $p_w$, for all $Re_\tau$ cases and for three different inner-scaled wall-normal positions. Note that the wavenumber representation on the axes includes a wall-scaling with $y$. The solid, coloured iso-contours correspond to the same two contour values as were considered in figure~\ref{fig:2Dcoh-log}($d$), $\gamma^2_{up_w} = [0.05;~0.25]$. ($b$,$d$,$f$) Similar to sub-figure ($a$) but now for $v$ and $p_w$, with the solid, coloured iso-contours corresponding to the same two contour values as were considered in figure~\ref{fig:2Dcoh-log}($e$). For indicating trends of scale-isotropy, dash-dotted arcs correspond to constant values of $\lambda/y = 1$, 3, 8, 14, 85, and 140 (from the most outer-arc going inward).}
\label{fig:2DcohRtrendlin}
\end{figure}

To present the 2D coherence for $p_w$ and $u$ as a function of $y$ and $Re_\tau$, we reside to a representation based on distance-from-the-wall scaling. First, all cases of $Re_\tau$ are shown for a height of $y^+ \approx 80$ in figure~\ref{fig:2DcohRtrendlin}($c$); the darkest red contour re-shows the contour of figure~\ref{fig:2Dcoh-log}($d$) for $Re_\tau \approx 5\,200$. Here the axes of the log-polar format are adapted to account for the wall-scaling: instead of $k^+_{\rm ref} = 1/50\,000$, the reference wavenumber is now made $y$-dependent following $k_{\rm yref} = 1/\left(1\,000y\right)$. This reference wavenumber is used in the new definition of the log-polar axes, $k_x^* = \xi k_x/k$ and $k_z^* = \xi k_z/k$, with $\xi = \log_{10}\left(k/k_{\rm yref}\right)$. Using the same format, two other wall-normal positions of $y^+ \approx 20$ (figure\ref{fig:2DcohRtrendlin}$a$) and $y^+ \approx 320$ (figure~\ref{fig:2DcohRtrendlin}$e$) are considered. Alongside, the 2D coherence for $p_w$ and $v$ is shown in the exact same format. All plots include six arcs of constant $ky$ (or constant $\lambda/y$ with the values stated in the caption), indicating trends of scale-isotropy. Lines of constant $k_z/k_x$ still correspond to the ones shown in figure~\ref{fig:2Dcoh-nor}, and indicate the $k_z = 2k_x$ trend (solid lines in figures~\ref{fig:2DcohRtrendlin}$a$-$f$) and $k_z = k_x$ trend (dashed lines in figures~\ref{fig:2DcohRtrendlin}$b$,$d$,$f$).
\begin{figure} 
\vspace{0pt}
\centering
\includegraphics[width = 0.999\textwidth]{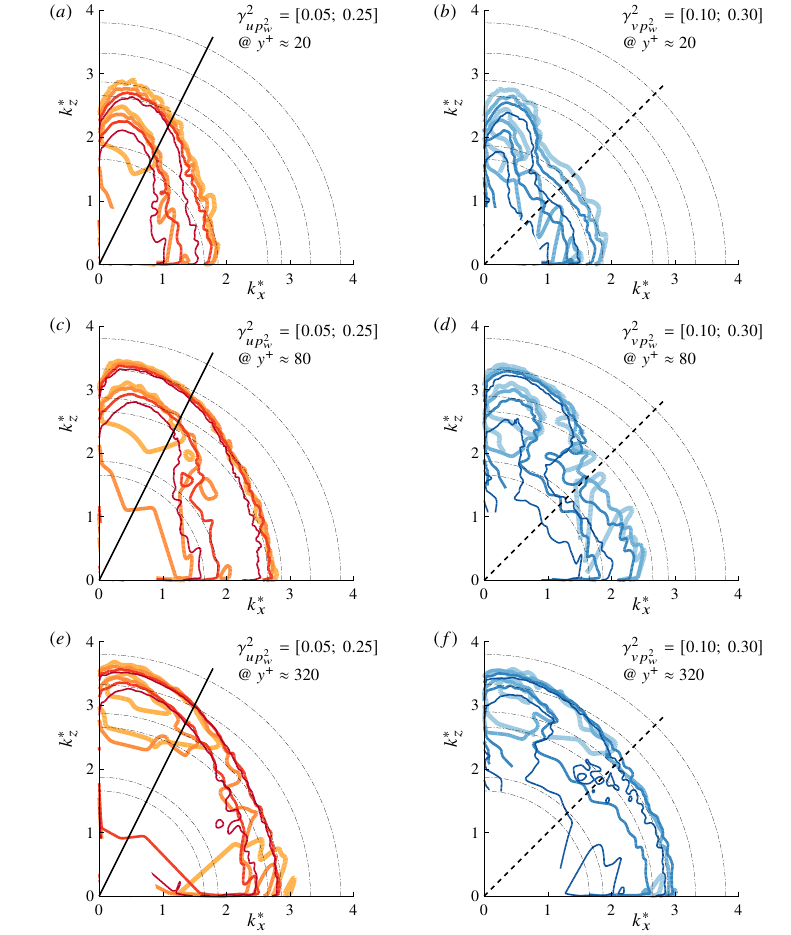}
\caption{Similar to figure~\ref{fig:2DcohRtrendlin}, but now for ($a$,$c$,$e$) $u$ and $p^2_w$ (wall-pressure squared) and ($b$,$d$,$f$) $v$ and $p^2_w$.}
\label{fig:2DcohRtrendnon}
\end{figure}

Iso-contours of $\gamma^2_{up_w}$ in figures~\ref{fig:2DcohRtrendlin}($a$,$c$,$e$) collapse well for all $Re_\tau$, in particular for the $\gamma^2_{up_w} = 0.05$ iso-contour. These plots reveal a wall-scaling of the 2D coherence because the iso-contours appear at the same position for all three $y$ positions, increasing consecutively by a factor of four ($y^+ \approx 20 \rightarrow 80 \rightarrow 320$). Note that iso-contours are less well bundled at $y^+ \approx 320$; this is ascribed to only the very large (less-converged) scales being coherent. Figures~\ref{fig:2DcohRtrendlin}($b$,$d$,$f$) present iso-contours of coherence for $p_w$ and $v$, and reveal a Reynolds number-invariant wall-scaling at the small-wavelength end. In contrast to $u$, the $v$ iso-contours of coherence adhere to trends of scale-isotropy for $p_wv$ structures that are not strongly elongated in $x$ (\emph{e.g.}, for scales residing at $k_z \lesssim 2k_x$). A maximum of the coherence resides near $\lambda/y = 8$ and is largely invariant with increasing $\lambda_x/\lambda_z$, except that the coherence shows a significant drop in amplitude for strongly-stretched structures in $x$ (this is more visible in figure~\ref{fig:2Dcoh-log}$f$). Finally, the wall-pressure is correlated stronger with $v$ than with $u$ ($\gamma^2_{vp_w} > \gamma^2_{up_w}$).

As for the 1D analysis in \S\,\ref{sec:1Dscaling}, the 2D coherence analysis is here extended to the coherence between the $u$ and $v$ motions and the wall-pressure squared. Contours of $\gamma^2_{up^2_w}$ and $\gamma^2_{vp^2_w}$ are shown in figures~\ref{fig:2DcohRtrendnon}($a$,$c$,$e$) and~\ref{fig:2DcohRtrendnon}($b$,$d$,$f$), respectively, and in a format identical to figure~\ref{fig:2DcohRtrendlin}. Coherence $\gamma^2_{up^2_w}$ does not obey a perfect wall-scaling, but for $y^+ \approx 80$ and $y^+ \approx 320$, the contours nearly collapse for all $Re_\tau$ and appear at similar locations in the plots. In general, the coherence with the wall-pressure-squared term resides at larger scales than the linear wall-pressure term, as was also the case in the 1D analysis. Again, iso-contours at the highest $y$ position correspond to much larger scales with less smooth spectral statistics, and for the two lowest Reynolds numbers, this position is beyond the logarithmic region. In the buffer region (at $y^+ \approx 20$ in figure~\ref{fig:2DcohRtrendnon}$a$), the coherence remains nearly Reynolds number-invariant and does no longer abide by a wall-scaling. This was also apparent from figure~\ref{fig:1Dcohnon}($a$). For the coherence of the wall-pressure-squared term with the $v$ motions, observations are similar.

\section{Exploring velocity-state estimation using sparse experimental data}\label{sec:state}
Now that Reynolds-number scalings of the wall-pressure--velocity were identified, experimental wall-pressure and velocity data are assessed for exploring whether velocity-state estimation based on wall-pressure input data is feasible (in practice). At first, the experimental findings are compared to the DNS-inferred correlations in \S\,\ref{sec:sparse}, after which the quadratic stochastic approach for velocity-state estimation is outlined in \S\,\ref{sec:SE}. Finally, in \S\,\ref{sec:bacc}, the accuracy in the velocity-estate estimates is discussed.
\begin{figure} 
\vspace{0pt}
\centering
\includegraphics[width = 0.999\textwidth]{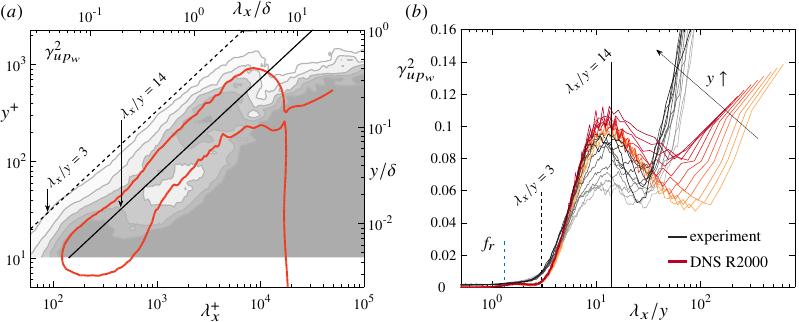}
\caption{($a$) Linear coherence spectrogram of $u$ and $p_w$, for the experimental data (grey scale filled iso-contours 0.02:0.02:0.1) and DNS R2000 data (a single red contour at $\gamma^2_{up_w} = 0.08$). ($b$) Linear coherence spectra for 11 positions within the logarithmic region, in the range $80 \lesssim y^+ \lesssim 0.15Re_\tau$. Results in frequency space are converted to a wavelength-dependence using $\lambda_x \equiv \overline{U}(y)/f$.}
\label{fig:explincoh}
\end{figure}

\subsection{Coherence of single-point input-output data}\label{sec:sparse}
With the available experimental time series of the streamwise velocity, $u(y,t)$, and the corresponding post-processed wall-pressure $p_w(t)$, the linear coherence $\gamma^2_{up_w}$ is computed in frequency space. Note that we only consider the wall-pressure--velocity correlation with the $u$ component since the single hot-wire probe measurements performed only provide streamwise velocity time series. Figure~\ref{fig:explincoh}($a$) shows the coherence spectrogram after conversion from a frequency-to-wavelength dependence, using the local mean velocity (\S\,\ref{sec:dataexp}). Overlaid is one iso-contour computed from the DNS R2000 data, highlighting that the ridge of strong coherence was well-captured by the experiment. Inspection of the slightly noisier experimental spectrogram is facilitated by showing individual coherence spectra in figure~\ref{fig:explincoh}($b$), for 11 positions within the logarithmic region taken as $80 \lesssim y^+ \lesssim 0.15Re_\tau$. Seemingly, the experimental spectra agree well in terms of the onset of coherence at a wall-scaling of around $\lambda_x/y = 3$ and its peak residing near $\lambda_x/y = 14$. For lower $y$ positions (a lower grey-scale intensity), the peak magnitude is underestimated compared to the DNS-inferred value of $\gamma^2_{up_w}\vert_{\rm peak} \approx 0.1$. A lower coherence is expected, given that denoising of experimental data is never perfect and thus leaves (incoherent) energy within the $p_w(t)$ signals.

A primary difference between the DNS and experimental coherence is observed for very large wavelengths. While coherence for the DNS increases beyond a constant $\lambda_x \approx 10\delta$ for all $y$, the experimental coherence shows a kick-up around $\lambda_x/y \approx 30$. We conjecture that the latter is related to the facility-noise filtering procedure of the experimental wall-pressure (described in Appendix~\ref{sec:appA}). That is, the energy content at larger streamwise wavelengths is attenuated in comparison to the DNS (see figure~\ref{fig:pwspectra}$b$ in Appendix~\ref{sec:appA}). This attenuated energy resides at relatively large \emph{spanwise scales} of $\lambda_z > 1.8\delta$, and these scales of the wall-pressure field are generally less coherent with the grazing velocity fluctuations. Hence, the retained energy in the noise-filtered experimental pressure data is relatively more coherent. Nevertheless, this section illustrates that with the adopted filtering procedures, the experiment is able to capture the general characteristics of the wall-pressure--velocity coupling. According to the authors' knowledge, this has not been attempted before and hence is an important result for the practical implementation of wall-pressure-based control.

\begin{figure} 
\vspace{0pt}
\centering
\includegraphics[width = 0.999\textwidth]{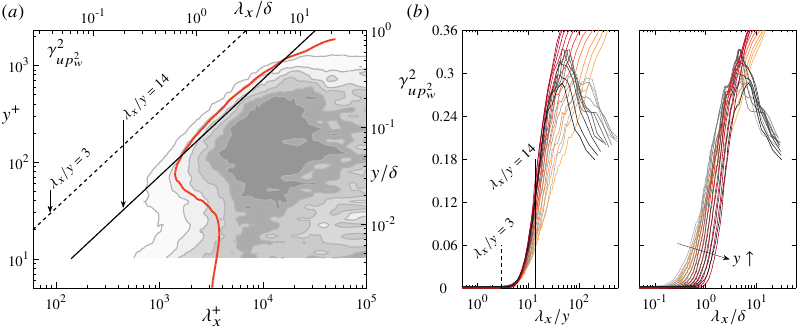}
\caption{($a$) Linear coherence spectrogram of $u$ and $p^2_w$, for the experimental data (grey scale filled iso-contours 0.04:0.04:0.24) and DNS R2000 data (a single red contour at $\gamma^2_{up^2_w} = 0.08$). ($b$) Linear coherence spectra for 11 positions of the experimental data within the logarithmic region, in the range $80 \lesssim y^+ \lesssim 0.15Re_\tau$. Results in frequency space are converted to a wavelength-dependence using $\lambda_x \equiv \overline{U}(y)/f$.}
\label{fig:expnoncoh}
\end{figure}

For the nonlinear pressure coherence, $\gamma^2_{up_w^2}$, the comparison between the experiment and DNS is presented in figure~\ref{fig:expnoncoh} in a similar way as for the linear coherence in figure~\ref{fig:explincoh}. Within the logarithmic region, the onset of coherence beyond $\lambda_x/y = 14$ is well captured. A slight discrepancy in the near-wall region between the experimental and DNS iso-contours of $\gamma^2_{up_w^2} = 0.08$ is ascribed to the difference in spatial and temporal coherence spectra. Recall the discussion of figure~\ref{fig:1Dcohnon} in that the coherence involving $p_w^2$ relates to an intensity modulation in which pressure fluctuations become more/less intense at a spatial scale that is a multiple of the dominant (linearly) correlated scale.

\subsection{Stochastic estimation}\label{sec:SE}
For a real-time controller working with wall-based sensing, a state estimation of the turbulence velocity fluctuations is critical. An overview of LSE and QSE was provided in \S\,\ref{sec:introscaling}, and these are here implemented in the time domain to illustrate the state estimation and quantify the estimation accuracy. Stochastic estimation procedures are performed in frequency space (not wavelength space) since the experimental data comprise time series. When concentrating on LSE, the time-domain estimate of the logarithmic region $u$ fluctuations at $y_e$ can be formed through the convolution:
\begin{equation}\label{eq:timeLSE}
    \widehat{u}_{\rm LSE}\left(y_e,t\right) = \left(h_l \circledast p_w\right)\left(t\right),
\end{equation}
where the temporal kernel is the inverse Fourier transform of the frequency-domain kernel, $h_l = \mathcal{F}^{-1}\left[H_L\right]$, with
\begin{equation}\label{eq:linkernel}
    H_L\left(y_e,f\right) = \frac{\phi_{up_w}\left(y_e;f\right)}{\phi_{p_wp_w}\left(f\right)} = \vert H_L\left(f\right)\vert \cdot e^{j \psi\left(y_e;f\right)}.
\end{equation}
The expression \eqref{eq:linkernel} is equivalent to \eqref{eq:introHL}, but for a 1D frequency-dependence, and the kernel's gain can be related to the coherence, according to
\begin{equation}\label{eq:kertocoh}
    \vert H_L\left(y_e;f\right) \vert = \sqrt{\gamma^2_{up_w}\left(y_e;f\right)\frac{\phi_{uu}\left(y_e;f\right)}{\phi_{p_wp_w}\left(f\right)}}.
\end{equation}
For QSE, the time-domain estimate can be written the first higher-order term of the input pressure $p_w$ \citep{naguib:2001a}:
\begin{equation}\label{eq:timeQSE}
    \widehat{u}_{\rm QSE}\left(y_e,t\right) = \left(h_l \circledast p_w\right)\left(t\right) + \left(h_q \circledast p^2_w\right)\left(t\right).
\end{equation}
As noted in \S\,\ref{sec:introscaling}, when the probability density function (PDF) of the input is symmetric, the linear kernel in the QSE is equal to the one in the LSE (for our experimental wall-pressure data the skewness is negligible, see figure~\ref{fig:pwpdf} in Appendix~\ref{sec:appA}). The quadratic kernel also follows from the inverse Fourier transform of the frequency-domain kernel, $h_q = \mathcal{F}^{-1}\left[H_Q\right]$, with 
\begin{equation}\label{eq:quakernel}
    H_Q\left(y_e,f\right) = \frac{\phi_{up^2_w}\left(y_e;f\right)}{\phi_{p^2_wp^2_w}\left(f\right)}.
\end{equation}
For completeness, the linear and quadratic kernels can also be written just in terms of the two-point correlations:
\begin{equation}\label{eq:timekernelsLSEQSE}
    h_l\left(y_e,\tau\right) = \frac{\langle p_w\left(t\right) u\left(y_e,t+\tau\right) \rangle}{\langle p^2_w\rangle};~~~~~~~~h_q\left(y_e,\tau\right) = \frac{\langle p^2_w\left(t\right) u\left(y_e,t+\tau\right) \rangle}{\langle p^2_w\rangle^2}.
\end{equation}

\begin{figure} 
\vspace{0pt}
\centering
\includegraphics[width = 0.999\textwidth]{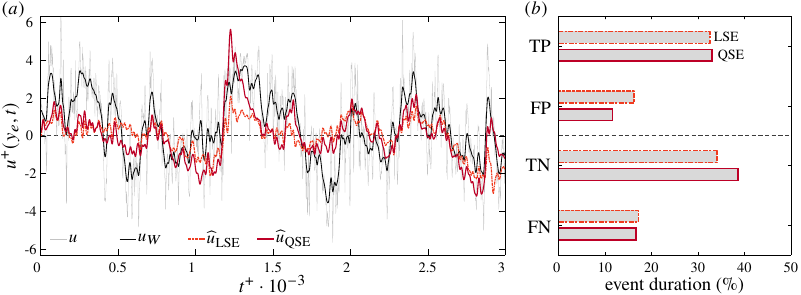}
\caption{($a$) Sample of the stochastic estimate of the $u$ fluctuations at $y_e^+ \approx 80$, based on wall-pressure, in comparison to a true (measured) time series. The measured hot-wire time series $u(y_e,t)$ is shown with a grey line, and is filtered to only retain the wall-attached $u$ fluctuations, $u_W(y_e,t)$, shown with the black line. Estimates using LSE, $\widehat{u}_{\rm LSE}(y_e,t)$, and QSE, $\widehat{u}_{\rm QSE}(y_e,t)$, are shown with orange and red lines, respectively. Correlation coefficients of the measured wall-attached $u$ fluctuations and the linear estimate ($u_W$ with $\widehat{u}_{\rm LSE}$), and the quadratic estimate ($u_W$ with $\widehat{u}_{\rm QSE}$), equal 0.48 and 0.60, respectively (see figure~\ref{fig:corrBACC}($b$) for $y_e^+ \approx 80$).} ($b$) Time-percentage of binary events in the LSE (orange dash-dotted) and QSE (red solid), relative to $u_W$.
\label{fig:timeseriesBACC}
\end{figure}

Estimates are now performed, for which transfer kernels were first generated from 50\,\% of the available data (thus from time series of $T_a/2$ long, still spanning more than 16\,000 boundary layer turnover times). The remaining time series data were used for the estimation. Results for an unconditional estimate of the $u$ fluctuations at $y^+_e \approx 80$ are shown in figure~\ref{fig:timeseriesBACC}($a$). Time series are shown for a total duration of $\Delta t^+ = 3000$, based on the LSE and QSE. Estimates are compared to the true (measured) time series: the raw measured time series is shown with the grey line, while a large-scaled filtered version, $u_W$, is shown with the black line. This latter signal $u_W$ only retains the wall-attached scales, which are defined as the streamwise velocity fluctuations that are correlated with the wall-shear stress (or friction velocity) fluctuations. Practically, $u_W$ is a large-scale pass-filtered signal of $u$, and the filter was derived from velocity-velocity correlations and was confirmed to be Reynolds-number invariant \citep{baars:2017a}. This spectral filter has a definitive cut-off at $\lambda_x/y = 14$ and thus also comprises a wall-scaling $\lambda_x \sim y$, meaning that only progressively larger scales are retained for larger $y$ positions. Further details of this filter can be found in the literature \citep[figure~9 and pp. 16-17 of][]{baars:2020P1a}. Note that $u_W$ will serve as a reference case for comparing the estimations to since the filtered fluctuations following a wall-scaling are more representative of the fluctuations that can physically be estimated using stochastic estimation. When inspecting figure~\ref{fig:timeseriesBACC}($a$), it is evident that the QSE procedure better estimates $u_W$ (and thus also $u$), but this is further quantified in the next section.

\subsection{Accuracy of velocity-state estimation}\label{sec:bacc}
We here assess the accuracy of the velocity-state estimates using a binary-state approach as well as a conventional correlation coefficient. When flow control systems would operate with on/off (binary) actuators only \citep{abbassi:2017a}, actuators would, in its simplest form, operate based on the estimated signal's sign (\emph{e.g.}, with a zero-valued threshold). In this context, the goodness of the estimates is quantified with a binary accuracy (BACC). When binarizing $u_W$ (here considered as the true signal) and the estimated signal ($\widehat{u}_{\rm LSE}$ or $\widehat{u}_{\rm QSE}$) at every time instant, only four events are possible: a true positive (TP) occurs when both signals are positive, whereas both signals being negative will yield a true negative (TN). Additionally, false positive (FP) and false negative (FN) linear estimates occur for $u_W(t) < 0$ and $\widehat{u}_{\rm LSE}(t) \geq 0$, or vice versa, respectively. The BACC defined as
\begin{equation}
    {\rm BACC} = \frac{T_{\rm TP}+T_{\rm TN}}{T},
\end{equation}
represents the cumulative time that the estimate is true positive and negative ($T_{\rm TP} + T_{\rm TN}$), relative to the total duration of the signal. Note that a BACC of unity does not mean that the estimate is perfect (that would be $\widehat{u}_{\rm LSE}(t) = u_W(t)$), but only that ${\rm sgn}\left[\widehat{u}_{\rm LSE}(t)\right] = {\rm sgn}\left[u_W(t)\right] \, \forall t$. Figure~\ref{fig:timeseriesBACC}($b$) presents the time-percentage of each of the four binary events for the full estimate of the case shown in figure~\ref{fig:timeseriesBACC}($a$). The total BACC for the LSE procedure is 66.7\,\%, while the QSE improves this to 71.7\,\%. This improvement comes from an increase of TN instances at the expense of less FP instances (such occasions appear around $t^+ \cdot 10^{-3} = 0.6$ and $t^+ \cdot 10^{-3} = 1.8$ in figure~\ref{fig:timeseriesBACC}$a$).

\begin{figure} 
\vspace{0pt}
\centering
\includegraphics[width = 0.999\textwidth]{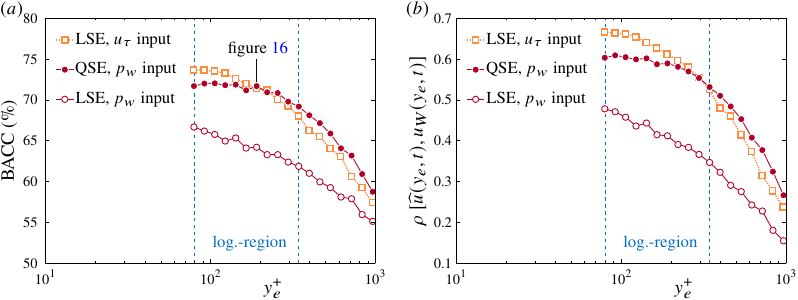}
\caption{($a$) BACC for a range of estimation positions, $y_e$. ($b$) Similar to sub-figure ($a$), but now for the correlation coefficient between the LSE- and QSE-based estimates and the true large-scales, taken as $u_W$.}
\label{fig:corrBACC}
\end{figure}

To quantify the goodness of the estimate further, the BACC is plotted for a range of $y_e$, starting at the lower end of the logarithmic region, $y_e^+ \approx 80$, up to the start of the intermittent region at $y_e^+ = 0.4Re_\tau$. Figure~\ref{fig:corrBACC}($a$) includes three profiles: the red ones are based on wall-pressure input, using both LSE and QSE, while the orange is based on an LSE with a representative wall-friction velocity input. This latter case considers data acquired in an identical experiment as the current one, except that the wall-pressure quantity was replaced with a single hot-film sensor on the wall, yielding a voltage signal as a surrogate for the fluctuations in friction velocity \citep[see][]{abbassi:2017a,dacome:2023a}. It is well-known that hot-film yields relatively clean signals (not subject to facility noise as is the wall-pressure) and that the linear correlation between the wall-signal and the off-the-wall velocity fluctuations is relatively strong \citep[\emph{e.g.},][]{hutchins:2011a,baars:2017a}. Note that it was confirmed that for this case with a friction-velocity input, a QSE does not result in an improved estimate per the findings of \citet{guezennec:1989a}. Finally, before interpreting the results, it is important to put the BACC magnitude into perspective. An estimate with a BACC of 50\,\% would reflect a random process and would thus be, on average, impractical when attempting real-time control based on such an estimate. In the real-time control work of \cite{abbassi:2017a} it was shown that real-time control with a BACC level of around 70\,\% was sufficient for targeting specific structures (\emph{e.g.}, positive or negative excursions in streamwise velocity). They used a friction-velocity input for their off-the-wall velocity state estimation, and their case is represented by the `LSE, $u_\tau$ input' curve in figure~\ref{fig:corrBACC}($a$). Hence, this case serves as a (successful) reference case.

For an LSE procedure with wall-pressure input, the BACC in the logarithmic region remains below 67\,\%. However, a QSE procedure results in a very similar BACC (mostly in excess of 72\,\%) as the reference case with the LSE based on a hot-film sensor-input. Hence, the current analysis shows that velocity-state estimation of the turbulent flow in the logarithmic region is viable with wall-based pressure sensing (provided that the quadratic pressure term is included), even when significant levels of facility noise are present.

\begin{figure} 
\vspace{0pt}
\centering
\includegraphics[width = 0.999\textwidth]{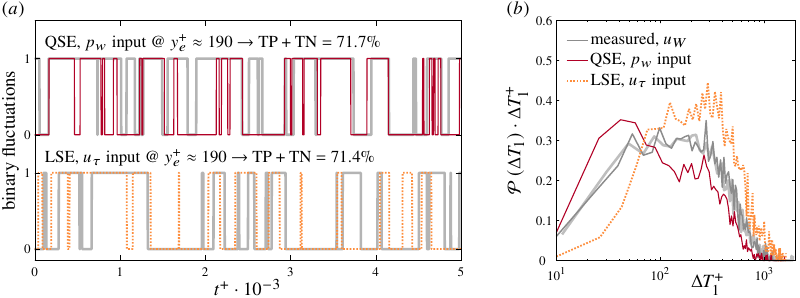}
\caption{($a$) Binary fluctuations of the QSE in solid dark red (top row) and LSE in dashed orange (bottom row) near the geometric centre of the logarithmic region, compared to the binary fluctuations of the true large-scales, $u_W$, in grey. ($b$) Premultiplied probability density functions of all uninterrupted time spans for which the binary signals equal unity ($\Delta T_1$).}
\label{fig:estdiff}
\end{figure}

Similar trends as for the BACC curves are found when concentrating on a conventional cross-correlation coefficient between the estimated signals and the true (large-scale) filtered signal $u_W$. Figure~\ref{fig:corrBACC}($b$) presents profiles of the correlation coefficient of the measured wall-attached $u$ fluctuations and the linear estimate ($u_W$ with $\widehat{u}_{\rm LSE}$), and the quadratic estimate ($u_W$ with $\widehat{u}_{\rm QSE}$). A correlation coefficient of $u_W$ with a wall-friction velocity signal is also shown for reference. As for the observations made based on the BACC, the inclusion of the quadratic term of the wall-pressure improves the correlation of the estimated signal with the true signal, $u_W$: the normalized correlation coefficient reaches 0.60 (at $y_e^+ = 80$) with the QSE, while the LSE only results in a correlation coefficient of 0.48. Thus, not only the BACC show a good promise for wall-based pressure sensing to perform off-the-wall velocity estimates, but further details of the velocity fluctuations themselves--as captured by a correlation coefficient--are also well-captured. In this regard, a final comparison between the binary fluctuations obtained through QSE with the wall-pressure, and LSE with the hot-film sensor-input is made in figure~\ref{fig:estdiff}. Near the geometric centre of the logarithmic region, at $y^+_e \approx 190$, the BACC and correlation coefficient are nearly identical (see figures~\ref{fig:corrBACC}$a$ and~\ref{fig:corrBACC}$b$). Binary fluctuations of the estimates, as well as the true large-scale signal $u_W$ are shown in figure~\ref{fig:estdiff}($a$). Note that these two sets of time series are not synchronized, since they come from different experiments. Differences in the estimations are summarized in terms of a PDF of all uninterrupted time spans for which the binary signals equal unity (denoted as time span $\Delta T_1$). This PDF is presented in premultiplied form (figure~\ref{fig:estdiff}$b$), so that longer time spans are weighted accordingly, \emph{e.g.}, the area-under-the-curve is representative of how much a range of $\Delta T_1$ contributes to the total time series-duration. PDFs corresponding to the true signals (the two grey binary fluctuations on the left) are equal as expected. Even though the BACC is equal for the two estimates, the binary fluctuations obtained with QSE ($p_w$ input) results in more `short' events, while the LSE ($u_\tau$ input) results in more `long' events (cross-over at $\Delta T^+_1 \approx 100$). So although the current findings show the feasibility of velocity-state estimation based on wall-pressure sensing, further research should address this accuracy in more detail by considering temporal-accuracy characteristics, its application in real-time control, and robustness to actuator noise and other external sources of noise.

\section{Concluding remarks}\label{sec:concl}
This work has been motivated by the potential feasibility of using wall-pressure input data for estimating off-the-wall turbulent velocities. Such an estimation capability is of high interest to realizing real-time controllers for wall-bounded turbulence---solely relying on wall pressure-input data---as long as they have a sufficient velocity-state estimation capability and are robust with variations in Reynolds number. The latter requires a firm understanding of the Reynolds-number scaling of wall-pressure--velocity correlations. In this regard, we examined the scaling behaviour of the coherence between the turbulent velocity fluctuations and the wall-pressure field, based on DNS data of turbulent channel flow with a Reynolds-number range $Re_\tau \approx 550$ -- $5\,200$. Several findings are summarized as follows:
\begin{enumerate}[labelwidth=0.85cm,labelindent=0pt,leftmargin=1.00cm,label=(\roman*),align=left]
\item \noindent Not only the 1D coherence analysis (\S\,\ref{sec:1Dscaling}), but also a 2D coherence analysis (\S\,\ref{sec:2Dscaling}) revealed that the coherence between $p_w$ and $u$, and $p_w$ and $v$, adhere to a Reynolds number-invariant wall-scaling with $y$.
\item \noindent A 1D streamwise data analysis with $u$ and $v$ velocity fluctuations at a given $y$ position revealed that they are most strongly coupled to the linear term of the wall-pressure fluctuations, $p_w$, at a scale of $\lambda_x/y \approx 14$ and $\lambda_x/y \approx 8$, respectively.
\item \noindent With the 2D extension of the analysis as function of $\lambda_x$ and $\lambda_z$, it was shown that the peak-coherence for $p_w$ and $u$ remains near $\lambda_x/y \approx 14$ and is reasonably symmetric around $\lambda_x/\lambda_z = 2$. The 2D coherence for $p_w$ and $v$ peaks around $\lambda_x/\lambda_z = 1.0$.
\item \noindent When statistically combining the results from the 1D and 2D analyses, it can be concluded that the ridge of coherence for $p_w$ and $u$ scales following $\lambda_x:\lambda_z:y \propto 14:7:1$, while for $p_w$ and $v$ the scaling is $\lambda_x:\lambda_z:y \propto 8:8:1.0$.
\item \noindent Based on the 2D analysis, turbulent motions in $v$ that are coherent with $p_w$ show a much stronger isotropic behaviour than the $u$ motions, and also possess a higher coherence magnitude. While the $u$ motions elongated in $x$ following $\lambda_x/\lambda_z \approx 2$ are most coherent, the $v$ motions that are most coherent reside near a 2D wave-vector of $\lambda/y = 8$ and is largely invariant with increasing $\lambda_x/\lambda_z$ up to only very strongly-stretched structures in $x$ beyond $\lambda_x \gtrsim 3\lambda_z$.
\item \noindent The coherence with the wall-pressure-squared term $p_w^2$ resides at larger scales than the linear wall-pressure term, as was also the case in the 1D analysis. The coherence involving $p_w^2$ is reminiscent of a (large-scale) intensity-modulation of the pressure fluctuations. This is because the coherence dominates at a spatial scale that is a multitude of the characteristic (linearly correlated) fluctuations and because the coherence with the quadratic term, $p_w^2$, is strongest for the velocity fluctuations taken in the logarithmic region. The latter reflects the general consensus that the modulation of wall-quantities is driven by energetic large-scale motions in the logarithmic region of the flow.
\end{enumerate}

With the DNS data spanning a decade in friction Reynolds number and the appearance of a clear Reynolds number invariance when adopting a wall-scaling $\lambda_x \sim y$, the current work suggests strong evidence that an extrapolation of the scaling laws to higher $Re_\tau$ conditions can be made. Moreover, an experimental dataset comprising simultaneous measurements of wall-pressure and velocity provided ample evidence, at one value of $Re_\tau \approx 2$k, that the DNS-inferred correlations can be replicated with experimental pressure data subject to significant levels of (acoustic) facility noise. It was furthermore shown that in order to reach similar levels of estimation accuracy in the wall-pressure based estimates, compared to estimates based on an input resembling friction velocity fluctuations, it is critical to include the quadratic pressure term. This is consistent with earlier observations of \citet{naguib:2001a} who explored a time-domain QSE for the estimates of the conditional streamwise velocity. An accuracy of up to 72\,\% in the binary state of the streamwise velocity fluctuations in the logarithmic region is achieved; this corresponds to a correlation coefficient of $\sim 0.6$. Since measuring the fluctuating wall-pressure is relatively robust in practice and is a viable quantity to measure on an aircraft fuselage, the current study is a step towards the implementation of a reliable \emph{flow state estimation} framework for wall-bounded turbulence based on wall-pressure.


\vspace{15pt}
\backsection[Acknowledgements]{We would like to give special thanks to Stefan Bernardy, Peter Duyndam and Frits Donker Duyvis for technical assistance during commencement of the experimental campaign.}

\backsection[Funding]{An award of computer time was provided by the Innovative and Novel Computational Impact on Theory and Experiment (INCITE) program. This research used resources of the Argonne Leadership Computing Facility, which is a DOE Office of Science User Facility supported under Contract DE-AC02-06CH11357. The authors also wish to gratefully acknowledge the Department of Flow Physics \& Technology of the Faculty of Aerospace Engineering at Delft University of Technology for financial support in establishing the experimental setup.}

\backsection[Declaration of interests]{The authors report no conflict of interest.}


\backsection[Author ORCIDs]{\\Woutijn J. Baars \hyperlink{https://orcid.org/0000-0001-2345-6789}{https://orcid.org/0000-0001-2345-6789}; \\Giulio Dacome \hyperlink{https://orcid.org/0009-0000-3088-2495}{https://orcid.org/0009-0000-3088-2495}; \\Myoungkyu Lee \hyperlink{https://orcid.org/0000-0002-5647-6265}{https://orcid.org/0000-0002-5647-6265}.}


\appendix
\section{\label{sec:appA} Post-processing of the experimental wall-pressure signals}
Wall-pressure measurements with microphones, surface-embedded behind pinholes, require several post-processing steps to yield valid time series of the fluctuating wall-pressure. Background noise is well-known to contaminate pressure measurements in wall-bounded turbulence, particularly at relatively low $Re_\tau$ when the facility noise can overshadow the turbulence-induced pressure fluctuations. For instance, \citet{klewicki:2008a} compiled an empirical relation for the inner-scaled wall-pressure intensity, as a function of $Re_\tau$,
\begin{equation}\label{eq:pklewicki}
    p^{\prime +}_{w} = \frac{p_{w}^{\prime}}{\tau_w} = \sqrt{6.5 + 2.30\ln\left(\frac{Re_\tau}{333}\right)}.
\end{equation}
According to this formulation, the expected wall-pressure intensity in the W-tunnel facility is $p_{w}^{\prime} \approx 1.13$\,Pa (with parameters of table~\ref{tab:data}). This equates to an overall sound pressure level of roughly $95$\,dB ($p_{\rm ref} = 20$\,$\upmu$Pa). For the non-anechoic facility, the noise level exceeds this value. That is, the acoustic pressure intensity in the potential flow was measured to be $105$\,dB, revealing the clear need for the implementation of a noise-cancelling scheme. In total, three corrections were implemented: \emph{step 1} uses the spanwise extent of the wall-pressure array (a unique feature of the current study), whereas steps 2 and 3 are conceptually similar to the procedure described in Appendix~A of \citet{tsuji:2007a} and \citet{gibeau:2021a}. These steps convert the measurement time series of the center pinhole-mounted microphone (further denoted as $p_{w1} = p_4$) to a corrected wall-pressure signal: $p_{w1} \rightarrow p_{w2} \rightarrow p_{w3} \rightarrow p_{w4}$. Representative positions corresponding to the first three signals are shown in figure~\ref{fig:HRkernel}($d$).\\
\begin{figure} 
\vspace{0pt}
\centering
\includegraphics[width = 0.999\textwidth]{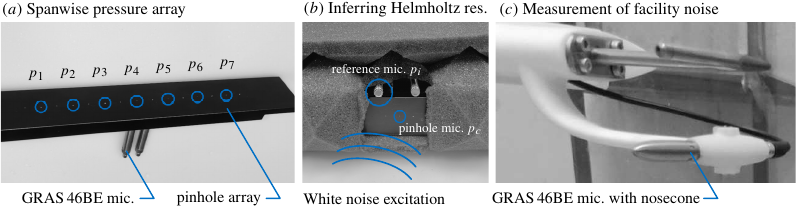}
\caption{($a$-$c$) Experimental hardware used to measure wall-pressure fluctuations in the W-tunnel facility.}
\label{fig:pwarray}
\end{figure}

\begin{figure} 
\vspace{0pt}
\centering
\includegraphics[width = 0.999\textwidth]{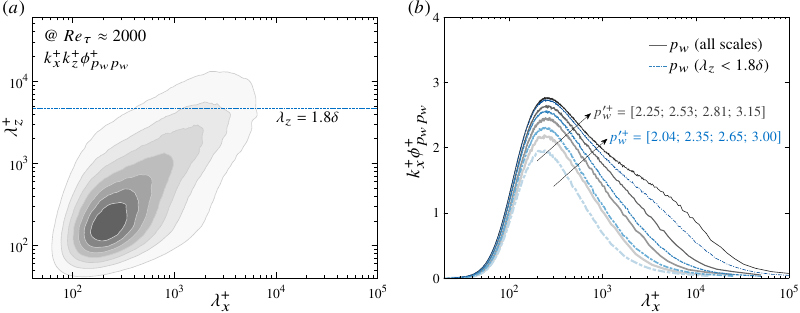}
\caption{($a$) 2D spectrogram of $p_w$ for the R2000 DNS case. ($b$) 1D wall-pressure spectra, constructed from the integration of the 2D spectra with the full range of spanwise scales (grey lines) and the spanwise scales limited to $\lambda_z < 1.8\delta$ (blue lines).}
\label{fig:2DpwDNS}
\end{figure}

\begin{figure} 
\vspace{0pt}
\centering
\includegraphics[width = 0.999\textwidth]{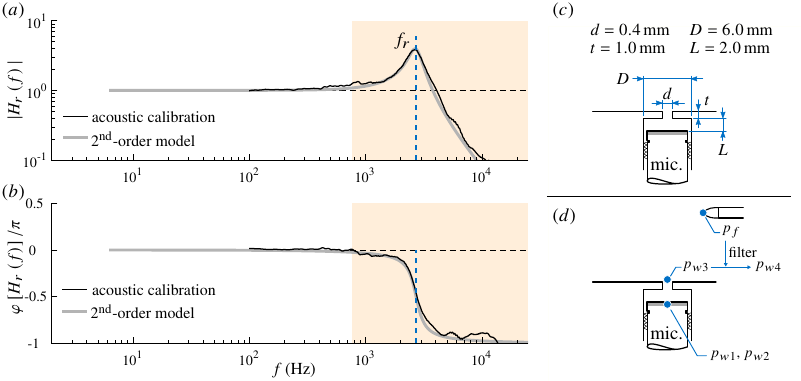}
\caption{($a$,$b$) Gain and phase of the resonator's transfer kernel. A comparison is made between the kernel obtained from the acoustic calibration, and a model kernel fitted to these empirical results. The filled area indicates the frequency range at which there is no expected coherence for the current study (further described in the text). ($b$) Geometric parameters of the pinhole-mounted microphone arrangement. ($c$) Wall-pressure signals that are being considered in the post-processing sequence.}
\label{fig:HRkernel}
\end{figure}

\noindent\textbf{Step 1: Subtract spanwise mean.} The spanwise array with $S = 7$ pinhole-mounted microphones spanned a total width of $\Delta z = 1.78\delta$, thus resolving spanwise wavelengths up to $\lambda_z \approx 1.8\delta$ (see figure~\ref{fig:pwarray}$a$). Note that the details of the sensors themselves (GRAS\,46BE $\nicefrac{1}{4}$\,in. microphones) and the pinhole-mounted microphone arrangement were described in \S\,\ref{sec:dataexp}. A spanwise-mean pressure at each temporal instant $t_i$, given by \eqref{eq:pwspan}, includes all energy in the pressure fluctuations residing at wavelengths $\lambda_z > 1.8\delta$.
\begin{equation}\label{eq:pwspan}
    p_{\rm span}\left(t_i\right) = \frac{1}{S}\sum_{s = 1}^{S} p_{s}\left(t_i\right)
\end{equation}
When subtracting this time series from measurement $p_{w1}$, a filtered signal follows,
\begin{equation}\label{eq:pw2}
    p_{w2}\left(t\right) = p_{w1}\left(t\right) - p_{\rm span}\left(t\right),
\end{equation}
and still corresponds to the position of the microphone diaphragm. Even though \eqref{eq:pw2} can be considered a conventional subtraction-based method \citep[as opposed to an optimal filtering-scheme, \emph{e.g.},][]{naguib:1996a}, the current scheme relies on the spanwise width and convergence of a 7-point mean. That is, all removed energy at $\lambda_z > 1.8\delta$ can only come from (1) the large-scale turbulence-induced wall-pressure, and/or (2) acoustic noise of the facility. Ideally, the latter is removed while the former is preserved. In order to assess how much energy of turbulence-induced pressure fluctuations resides at those scales, the DNS data are used. A 2D wall-pressure spectrum at $Re_\tau \approx 2000$ is shown in figure~\ref{fig:2DpwDNS}($a$), together with the spanwise cutoff wavelength of the pressure-array. When integrating over $\lambda_z$, the 1D streamwise spectrum is obtained and shown in figure~\ref{fig:2DpwDNS}($b$). In fact, two sets of spectra are shown: the raw 1D spectra and the ones that exclude the energy at $\lambda_z > 1.8\delta$ for all four Reynolds numbers (but note that only the R2000 case is relevant for the current experiments at closely matched $Re_\tau$). It is evident that only a small degree of large-scale energy is removed by subtracting the spanwise-mean pressure over the array, as evidenced by the integrated wall-pressure intensities listed in the annotations. For the R2000 case, the difference in integrated spectral-energy is $\approx 6$\,\%.

\begin{figure} 
\vspace{0pt}
\centering
\includegraphics[width = 0.999\textwidth]{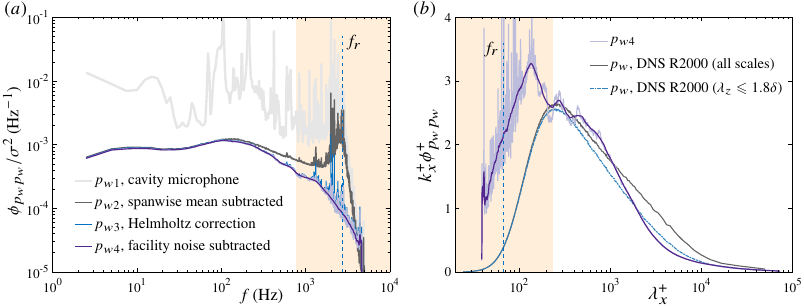}
\caption{($a$) Spectra of the various wall-pressure signals generated in the post-processing procedure. ($b$) Spectrum of the final, corrected wall-pressure signal obtained from experiments, compared to the spatial wall-pressure spectra from DNS.}
\label{fig:pwspectra}
\end{figure}

Since facility noise generally propagates as plane waves, acoustic noise reaches all microphones at equal times. Hence, subtraction of the spanwise mean through \eqref{eq:pw2} is an effective denoising procedure as seen from the spectra of the raw ($p_{w1}$) and spanwise-filtered ($p_{w2}$) pressure signals in figure~\ref{fig:pwspectra}($a$).\\

\noindent\textbf{Step 2: Helmholtz correction.} A pinhole configuration to enhance the spatial resolution of the microphone results in an unavoidable sub-surface cavity above the diaphragm, and thus Helmholtz resonance. Resonance amplifies the measured cavity pressure, relative to the excitation signal at the pinhole-inlet, at frequencies close to resonance. When sizing the resonator, the design resonance frequency, $f_0$, was computed from the formulation \citep{ingard:1953a,panton:1975b},
\begin{equation}\label{eq:f0}
    f_0 = \frac{a_0}{2\pi} \sqrt{\frac{S}{V(t+t^*) + L^2S/3}},
\end{equation}
where $a_0$ is the speed of sound, $S = \pi d_0^2/4$ is the area of the orifice, $V = \pi D^2 L/4$ is the cavity volume, and $t^*$ is an `end correction'. This correction is required, as besides the air within the neck a small portion of air inside the cavity and outside of the orifice will displace as well. When taking the end correction from \citet{ingard:1953a}, $t^* = 0.48\sqrt{S}\left(1-1.25d_0/D\right)$ and valid for $d_0/D < 0.4$ only, the design resonance frequency becomes $f_0 = 2\,725$\,Hz (dimensions of the pinhole and cavity are shown in figure~\ref{fig:HRkernel}$c$). Amplification of energy surrounding the resonance frequency is apparent from the broadband hump in the spectrum of $p_{w2}$ in figure~\ref{fig:pwspectra}($a$).

Correcting for the Helmholtz resonance is straightforward once the resonator has been characterized. Whereas \eqref{eq:f0} provides a design frequency $f_0$, it does not consider how the amplitude and phase of the measured pressure in the cavity relates to the inlet pressure. This can be modelled through a single input/output impulse transfer function, $H_r($f$)$, by considering the mass-spring-damper system-equivalent of the resonator. Its gain and phase are formulated as
\begin{equation}\label{eq:HR2}
    \vert H_r\left(f\right)\vert = \left[\left(1-\left(\frac{f}{f_r}\right)^2\right)^2 + \left(\frac{2\xi f}{f_r}\right)^2\right]^{-0.5},~\text{and}~\varphi\left[H_r\left(f\right) \right] = -\tan^{-1}\left[\frac{2\xi\left(f/f_r\right)}{1-\left(f/f_r\right)^2}\right].
\end{equation}
Here, $\xi$ is a damping constant and $f_r$ is the resonance frequency. Both $\xi$ and $f_0$ require empirical characterization through a calibration experiment, using acoustic pressure excitation. This experiment was conducted in a facility that is anechoic at frequencies above 200\,Hz \citep{merino:2020a}. White noise was produced with a Bose\textsuperscript{\textregistered} speaker and was recorded simultaneously by a reference microphone (measuring the excitation pressure at the inlet, $p_i$) and the cavity microphone ($p_c$), see figure\,\ref{fig:pwarray}($b$). Signals were acquired for a length of $T_a = 150$\,seconds at a sample rate of $f_s = 51.2$\,kHz. With the signal recordings, the experimental characterization of the Helmholtz resonator can be obtained through dividing the input-output cross-spectrum by the input spectrum: 
\begin{equation}\label{eq:HRkernel}
    \vert H_r^{\rm exp}\vert = \frac{\langle P_c\left(f\right)P_i\left(f\right)\rangle}{\langle P_i\left(f\right) P^*_i\left(f\right)\rangle}.
\end{equation}
Ensemble-averaging was conducted using FFT partitions of $N = 2^{13}$ samples, resulting in a spectral resolution of $df = 6.25$\,Hz and 1870 ensembles with 50\% overlap. The resultant gain and phase of the complex-valued kernel are plotted in figures~\ref{fig:HRkernel}($a$,$b$), but only for $f > 100$\,Hz due to the non-anechoic nature of the facility at lower frequencies, causing non-physical behaviour. The 2$^{\rm nd}$-order model transfer function $H_r($f$)$ was fit to the data to infer the resonance frequency and damping coefficient, yielding $f_r = 2\,725$\,Hz and $\xi = 0.128$, respectively. The measured kernel captures the experimental behaviour well, and the resonance frequency closely resembles the design frequency ($f_r/f_0 \approx 0.99$). The gain function shows the region of amplification around $f_r$, and the resonance peak is associated with the well-known phase-reversal between the excitation signal and cavity response.

Note that the time series of pressure at the pinhole-inlet can be generated by a convolution of the cavity pressure signal with the time-domain transfer kernel,
\begin{equation}\label{eq:pw3}
    p_{w3}\left(t\right) = \left(h_r^{\rm exp} \circledast p_{w2}\right)\left(t\right).
\end{equation}
Here $h_r^{\rm exp} = \mathcal{F}^{-1}\left[H_r^{\rm exp}\right]$ is taken from the experimental characterization. Before proceeding, it is of important to provide a note on the frequency range of interest. To utilize the wall-pressure for estimating the state of the off-the-wall $u$-velocity, the highest frequency of interest is related to the smallest coherent scales between $p_w$ and $u$. It was discussed in \S\,\ref{sec:1Dscaling} and \S\,\ref{sec:sparse} that only $u$-velocity scales at $\lambda_x/y \gtrsim 3$ are coherent with the wall-pressure. For a start of the logarithmic region at $y^+ = 80$, this corresponds to scales of $\lambda_x^+ \leq 240$ and a frequency range below $f = \overline{U}_c/\lambda_x \approx 800$\,Hz (convection velocity taken as the mean velocity at $y^+ = 80$). Hence, the frequencies of interest reside at frequencies on the left of the hatched region in figures~\ref{fig:HRkernel}($a$,$b$). The gain and phase are relatively flat for this region, and hence the Helmholtz correction has a minimal impact on the pressure fluctuations in this range. Nevertheless, the correction is still required for the validity of the wall-pressure spectrum at smaller scales. In this regard, the correction following \eqref{eq:pw3} can amplify high-frequency noise due to the fast decay of the gain beyond $f_r$. An application of a low-pass filter prevents this and was implemented following the literature \citep{tsuji:2007a,gibeau:2021a} with a cutoff frequency of $f\nu/u_\tau^2 = 0.25$ ($f = 3$\,kHz). A spectrum of the Helmholtz-corrected pressure fluctuations $p_{w3}$ is shown in figure~\ref{fig:pwspectra}($a$) and shows the removal of the amplified energy near $f_r$.\\

\noindent\textbf{Step 3: Remove facility noise.} A final step involves the removal of any remaining facility noise with the aid of the free-stream acoustic measurement, $p_f(t)$. This measurement was achieved by mounting a microphone in the potential flow region. A GRAS\,RA0022 $\nicefrac{1}{4}$\,in. the nosecone was installed to remove as much as possible the pressure fluctuations from the turbulence in the stagnation point (see photograph in figure~\ref{fig:pwarray}$c$). The noise removal procedure was implemented according to the description in \citet{gibeau:2021a} and is only briefly summarized here. A subtraction of an estimate of the facility noise, denoted as $\widehat{p}_{f}$, from the wall-pressure signal at the inlet of the pinhole-inlet, $p_{w3}$, is implemented following
\begin{equation}\label{eq:pw4}
    p_{w4}\left(t\right) = p_{w3}\left(t\right) - \widehat{p}_{f}\left(t\right).
\end{equation}
The estimate of the facility noise does not equal the measured facility noise, $p_f$, given that the noise is measured at a different position (in the potential flow region). Moreover, even though a nosecone was installed, the measurement is still intrusive and can be subject to self-induced pressure fluctuations. To generate the estimate, the Wiener noise cancelling filter coefficients can be derived from the measurement data and be implemented through the convolution of a digital FIR filter: $\widehat{p}_{f}\left(t_d\right) = \left(c \circledast p_f\right)(t_d)$. To emphasize the discrete-time dependence, subscript `d' is used. Filter coefficients $c(t_d)$ are defined for $m$ time instants and they are obtained through the Wiener-Hopf equations, $\mathbf{R}c = r$, in which $R$ is a symmetric Toeplitz-matrix of size $m \times m$ with the auto-correlations of $p_f$, and $r$ is the two-point cross-correlation vector of $p_{w3}$ and $p_f$ and has size $m \times 1$. The unique solution $c = \mathbf{R}^{-1}r$ yields the filter coefficients. The filter step is implemented with an order of $m = 30 \cdot 10^3$ on the signals that were down-sampled by a factor of 5, thus leaving an \emph{effective} sampling frequency of $f_s = 51.2/5 = 10.24$\,kHz. A spectrum of the pressure fluctuations after the removal of the facility noise through \eqref{eq:pw4} is shown in figure~\ref{fig:pwspectra}($a$) and highlights the removal of noise peaks that were still present in the $p_{w3}$ signal. Finally, it was also attempted to directly remove all tunnel noise from only the center pinhole-mounted microphone (without first performing \emph{step 1}), however, that resulted in a noisier final spectrum.

The spectrum of $p_{w4}$ is compared to the 1D spatial spectrum of DNS in figure~\ref{fig:pwspectra}($b$). Here frequency was transformed to wavelength using $\lambda_x = \overline{U}_c/f$ and a convection velocity of $U^+_c = 10$. In terms of integrated energy, the inner-scaled wall-pressure intensity from the experiments, $p^{\prime +}_{w4} \approx 3.04$, compares well with \eqref{eq:pklewicki} \citep{farabee:1991a,klewicki:2008a} predicting a value of 3.31. The spectral density itself compares reasonably well at the mid-scale range. At the large wavelength-end of the spectrum the attenuation is expected due to the over-excessive removal of energy in \emph{step 1}. At the small wavelength-end, the energy is amplified instead, presumably due to shortcomings in removing the Helmholtz resonance. That is, the transfer function in \emph{step 2} was generated using an acoustic calibration. However, a resonator behaves differently when excited by a grazing TBL flow, in comparison to an acoustic wave-excitation, since the `end correction' is changed and thus the resonance frequency and effective damping \citep{panton:1975b}. Nevertheless, the mismatch of the spectral shape does not affect any conclusions of the current work, given that the resonance is outside of the frequency range at which the scales of $u$ and $p_{w4}$ possess a non-zero coherence. Finally, comparing the experimental spectrum based on temporal data with the spatial DNS spectra is challenging. Frequency-to-wavenumber conversions are non-trivial and Taylor’s hypothesis can introduce aliasing-type discrepancies, particularly in the near-wall region \citep{perry:1990a,delalamo:2009a}.

\begin{figure} 
\vspace{0pt}
\centering
\includegraphics[width = 0.999\textwidth]{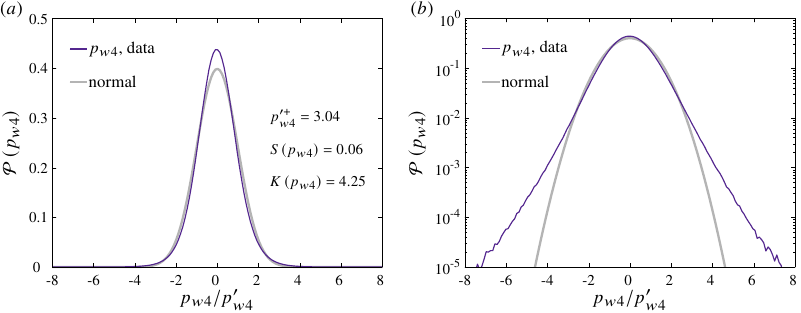}
\caption{Probability density functions of the fluctuating wall-pressure signal, $p_{w4}$, with a ($a$) linear axis and ($b$) logarithmic axis. A normal PDF is shown for reference.}
\label{fig:pwpdf}
\end{figure}

With the valid time series of the fluctuating wall-pressure, $p_{w4}$, the probability density function (PDF) can be generated. Figure~\ref{fig:pwpdf}($a$) shows the PDF with a linear ordinate and a normalization by the standard deviation. Around the zero-amplitude fluctuations, the PDF is larger than the associated normal distribution. In the tails of the PDF, being more evident with the semi-logarithmic axes in figure~\ref{fig:pwpdf}($b$), a larger amplitude is also noticeable and the PDF tends more towards an exponential form. Similar PDF's were shown in the literature \citep{tsuji:2007a,klewicki:2008a}. \citet{tsuji:2007a} indicated that the skewness increased with the Reynolds number, from -0.05 to 0.09 for $Re_\theta$ ranging from 5\,870 to 16\,700. Our skewness of 0.06 at $Re_\theta \approx 6\,190$ is slightly larger, but the exact skewness value is known to also depend on the value of $d^+$ \citep{schewe:1983a,gravante:1998a}. Nevertheless, skewness values are small, and third-order moments in the QSE approach can therefore be neglected \citep{naguib:2001a}.

\bibliographystyle{jfm}
\bibliography{bibtex_database}

\begin{thebibliography}{71}
\expandafter\ifx\csname natexlab\endcsname\relax\def\natexlab#1{#1}\fi
\def\au#1{#1} \def\ed#1{#1} \def\yr#1{#1}\def\at#1{#1}\def\jt#1{\textit{#1}}
  \def\bt#1{#1}\def\bvol#1{\textbf{#1}} \def\vol#1{#1} \def\pg#1{#1}
  \def\publ#1{#1}\def\arxiv#1{#1}\def\org#1{#1}\def\st#1{\textit{#1}}

\bibitem[Abbassi {\em et~al.\/}(2017)Abbassi, Baars, Hutchins \&
  Marusic]{abbassi:2017a}
{\sc \au{Abbassi, M.~R.}, \au{Baars, W.~J.}, \au{Hutchins, N.} \& \au{Marusic,
  I.}} \yr{2017}  \at{Skin-friction drag reduction in a high-{R}eynolds-number
  turbulent boundary layer via real-time control of large-scale structures}.
  \jt{Int. J. Heat Fluid Flow}  \bvol{67},  \pg{30--41}.

\bibitem[Adrian(1979)]{adrian:1979a}
{\sc \au{Adrian, R.~J.}} \yr{1979}  \at{Conditional eddies in isotropic
  turbulence}.  \jt{Phys. Fluids}  \bvol{22}~(11),  \pg{2065--2070}.

\bibitem[Adrian \& Moin(1988)]{adrian:1988a}
{\sc \au{Adrian, R.~J.} \& \au{Moin, P.}} \yr{1988}  \at{Stochastic estimation
  of organized turbulent structure: homogeneous shear flow}.  \jt{J. Fluid.
  Mech.}  \bvol{190},  \pg{531--559}.

\bibitem[Adrian {\em et~al.\/}(1987)Adrian, Moin \& Moser]{adrian:1987a}
{\sc \au{Adrian, R.~J.}, \au{Moin, P.} \& \au{Moser, R.~D.}} \yr{1987}
  \at{Stochastic estimation of conditional eddies in turbulent channel flow}.
  \jt{Proc. of the Summer Program 1987, Center for Turbulence Research}
  \pg{pp. 7--19}, {C}{T}{R}-S87.

\bibitem[Arun {\em et~al.\/}(2023)Arun, Bae \& {McKeon}]{arun:2023a}
{\sc \au{Arun, R.}, \au{Bae, H.~J.} \& \au{{McKeon}, B.~J.}} \yr{2023}
  \at{Towards real-time reconstruction of velocity fluctuations in turbulent
  channel flow}.  \jt{Phys. Rev. Fluids}  \bvol{8}, 064612.

\bibitem[Baars {\em et~al.\/}(2017)Baars, Hutchins \& Marusic]{baars:2017a}
{\sc \au{Baars, W.~J.}, \au{Hutchins, N.} \& \au{Marusic, I.}} \yr{2017}
  \at{Self-similarity of wall-attached turbulence in boundary layers}.  \jt{J.
  Fluid Mech.}  \bvol{823},  \pg{R2}.

\bibitem[Baars \& Marusic(2020)]{baars:2020P1a}
{\sc \au{Baars, W.~J.} \& \au{Marusic, I.}} \yr{2020}  \at{Data-driven
  decomposition of the streamwise turbulence kinetic energy in boundary layers.
  {P}art 1. {E}nergy spectra}.  \jt{J. Fluid Mech.}  \bvol{882},  \pg{A25}.

\bibitem[Baars \& Tinney(2014)]{baars:2014a}
{\sc \au{Baars, W.~J.} \& \au{Tinney, C.~E.}} \yr{2014}  \at{Proper orthogonal
  decomposition-based spectral higher-order stochastic estimation}.  \jt{Phys.
  Fluids}  \bvol{26}, 055112.

\bibitem[Bai {\em et~al.\/}(2014)Bai, Zhou, Zhang, Xu, Wang \&
  Antonia]{bai:2014a}
{\sc \au{Bai, H.~L.}, \au{Zhou, Y.}, \au{Zhang, W.~G.}, \au{Xu, S.~J.},
  \au{Wang, Y.} \& \au{Antonia, R.~A.}} \yr{2014}  \at{Active control of a
  turbulent boundary layer based on local surface perturbation}.  \jt{J. Fluid
  Mech.}  \bvol{750},  \pg{316--354}.

\bibitem[Baidya {\em et~al.\/}(2017)Baidya, Philip, Hutchins, Monty \&
  Marusic]{baidya:2017a}
{\sc \au{Baidya, R.}, \au{Philip, J.}, \au{Hutchins, N.}, \au{Monty, J.~P.} \&
  \au{Marusic, I.}} \yr{2017}  \at{Distance-from-the-wall scaling of turbulent
  motions in wall-bounded flows}.  \jt{Phys. Fluids}  \bvol{29}, 020712.

\bibitem[Canton {\em et~al.\/}(2016)Canton, \"{O}rlu, Chin \&
  Schlatter]{canton:2016a}
{\sc \au{Canton, J.}, \au{\"{O}rlu, R.}, \au{Chin, C.} \& \au{Schlatter, P.}}
  \yr{2016}  \at{Reynolds number dependence of large-scale friction control in
  turbulent channel flow}.  \jt{Phys. Rev. Fluids}  \bvol{1}, 081501.

\bibitem[Chauhan {\em et~al.\/}(2009)Chauhan, Monkewitz \&
  Nagib]{chauhan:2009a}
{\sc \au{Chauhan, K.~A.}, \au{Monkewitz, P.~A.} \& \au{Nagib, H.~M.}} \yr{2009}
   \at{Criteria for assessing experiments in zero pressure gradient boundary
  layers}.  \jt{Fluid Dyn. Res.}  \bvol{41}, 021404.

\bibitem[Chernyshenko(2021)]{chernyshenko:2021a}
{\sc \au{Chernyshenko, S.}} \yr{2021}  \at{{Extension of QSQH theory of scale
  interaction in near-wall turbulence to all velocity components}}.  \jt{J.
  Fluid Mech.}  \bvol{916},  \pg{A52}.

\bibitem[Choi {\em et~al.\/}(1994)Choi, Moin \& Kim]{choi:1994a}
{\sc \au{Choi, H.}, \au{Moin, P.} \& \au{Kim, J.}} \yr{1994}  \at{Active
  turbulence control for drag reduction in wall-bounded flows}.  \jt{J. Fluid
  Mech.}  \bvol{262},  \pg{75--110}.

\bibitem[Choi {\em et~al.\/}(1998)Choi, De{B}isschop \& Clayton]{choi:1998a}
{\sc \au{Choi, {K.-S.}}, \au{De{B}isschop, {J.-R.}} \& \au{Clayton, B.~R.}}
  \yr{1998}  \at{Turbulent boundary-layer control by means of spanwise-wall
  oscillation}.  \jt{AIAA J.}  \bvol{36}~(7),  \pg{1157--1163}.

\bibitem[Choi {\em et~al.\/}(2011)Choi, Jukes \& Whalley]{choi:2011a}
{\sc \au{Choi, {K.-S.}}, \au{Jukes, T.} \& \au{Whalley, R.}} \yr{2011}
  \at{Turbulent boundary-layer control with plasma actuators}.  \jt{Phil.
  Trans. R. Soc. A}  \bvol{369},  \pg{1443--1458}.

\bibitem[Cui \& Jacobi(2021)]{cui:2021a}
{\sc \au{Cui, G.} \& \au{Jacobi, I.}} \yr{2021}  \at{Biphase as a diagnostic
  for scale interactions in wall-bounded turbulence}.  \jt{Phys. Rev. Fluids}
  \bvol{6}, 014604.

\bibitem[Dacome {\em et~al.\/}(2023)Dacome, M\"{o}rsch, Kotsonis \&
  Baars]{dacome:2023a}
{\sc \au{Dacome, G.}, \au{M\"{o}rsch, R.}, \au{Kotsonis, M.} \& \au{Baars,
  W.~J.}} \yr{2023}  \at{Opposition flow control for reducing skin-friction
  drag of a turbulent boundary layer}.  \jt{arXiv:} 10.48550/arXiv.2311.00000.

\bibitem[{del \'{A}lamo} \& Jim\'{e}nez(2003)]{delalamo:2003a}
{\sc \au{{del \'{A}lamo}, J.~C.} \& \au{Jim\'{e}nez, J.}} \yr{2003}
  \at{Spectra of the very large anisotropic scales in turbulent channels}.
  \jt{Phys. Fluids}  \bvol{15}~(6),  \pg{L41--L44}.

\bibitem[{del \'{A}lamo} \& Jim\'{e}nez(2009)]{delalamo:2009a}
{\sc \au{{del \'{A}lamo}, J.~C.} \& \au{Jim\'{e}nez, J.}} \yr{2009}
  \at{{Estimation of turbulent convection velocites and corrections to Taylor's
  approximation}}.  \jt{J. Fluid Mech.}  \bvol{640},  \pg{5--26}.

\bibitem[Deng {\em et~al.\/}(2016)Deng, Huang \& Xu]{deng:2016a}
{\sc \au{Deng, B.~Q.}, \au{Huang, W.~X.} \& \au{Xu, C.~X.}} \yr{2016}
  \at{{Origin of effectiveness degradation in active drag reduction control of
  turbulent channel flow at $Re_\tau = 1000$}}.  \jt{J. Turbul.}
  \bvol{17}~(8),  \pg{758--786}.

\bibitem[Deshpande {\em et~al.\/}(2020)Deshpande, Chandran, Monty \&
  Marusic]{deshpande:2020a}
{\sc \au{Deshpande, R.}, \au{Chandran, D.}, \au{Monty, J.~P.} \& \au{Marusic,
  I.}} \yr{2020}  \at{Two-dimensional cross-spectrum of the streamwise velocity
  in turbulent boundary layers}.  \jt{J. Fluid Mech.}  \bvol{890},  \pg{R2}.

\bibitem[Ewing \& Citriniti(1999)]{ewing:1999c}
{\sc \au{Ewing, D.} \& \au{Citriniti, J.}} \yr{1999} Examination of a
  {L}{S}{E}/{P}{O}{D} complementary technique using single and multi-time
  information in the axisymmetric shear layer.  \bt{In {\em Proceedings of the
  IUTAM Symposium on simulation and identification of organized structures in
  flows\/}},  \pg{pp. 375--384}.  \publ{Lynby, Denmark: IUTAM}, eds. J. N.
  Sorensen, E. J. Hopfinger \& N. Aubry.

\bibitem[Farabee \& Casarella(1991)]{farabee:1991a}
{\sc \au{Farabee, T.~M.} \& \au{Casarella, M.~J.}} \yr{1991}  \at{Spectral
  features of wall pressure fluctuations beneath turbulent boundary layers}.
  \jt{Phys. Fluids A}  \bvol{3}~(10),  \pg{2410--2420}.

\bibitem[Ghaemi \& Scarano(2013)]{ghaemi:2013a}
{\sc \au{Ghaemi, S.} \& \au{Scarano, F.}} \yr{2013}  \at{Turbulent structure of
  high-amplitude pressure peaks within the turbulent boundary layer}.  \jt{J.
  Fluid Mech.}  \bvol{735},  \pg{381--426}.

\bibitem[Gibeau \& Ghaemi(2021)]{gibeau:2021a}
{\sc \au{Gibeau, B.} \& \au{Ghaemi, S.}} \yr{2021}  \at{Low- and mid-frequency
  wall-pressure sources in a turbulent boundary layer}.  \jt{J. Fluid Mech.}
  \bvol{918},  \pg{A18}.

\bibitem[Gravante {\em et~al.\/}(1998)Gravante, Naguib, Wark \&
  Nagib]{gravante:1998a}
{\sc \au{Gravante, S.~P.}, \au{Naguib, A.~M.}, \au{Wark, C.~E.} \& \au{Nagib,
  H.~M.}} \yr{1998}  \at{Characterization of the pressure fluctuations under a
  fully developed turbulent boundary layer}.  \jt{AIAA J.}  \bvol{36}~(10),
  \pg{1808--1816}.

\bibitem[Guastoni {\em et~al.\/}(2021)Guastoni, G\:{u}mes, Ianiro, Discetti,
  Schlatter, Azizpour \& Vinuesa]{guastoni:2021a}
{\sc \au{Guastoni, L.}, \au{G\:{u}mes, A.}, \au{Ianiro, A.}, \au{Discetti, S.},
  \au{Schlatter, P.}, \au{Azizpour, H.} \& \au{Vinuesa, R.}} \yr{2021}
  \at{Convolutional-network models to predict wall-bounded turbulence from wall
  quantities}.  \jt{J. Fluid Mech.}  \bvol{928},  \pg{A27}.

\bibitem[Guezennec(1989)]{guezennec:1989a}
{\sc \au{Guezennec, Y.~G.}} \yr{1989}  \at{Stochastic estimation of coherent
  structures in turbulent boundary layers}.  \jt{Phys. Fluids A}  \bvol{1}~(6),
   \pg{1054--1060}.

\bibitem[Hamilton {\em et~al.\/}(1995)Hamilton, Kim \& Waleffe]{hamilton:1995a}
{\sc \au{Hamilton, J.~M.}, \au{Kim, J.} \& \au{Waleffe, F.}} \yr{1995}
  \at{Regeneration mechanisms of near-wall turbulence structures}.  \jt{J.
  Fluid Mech.}  \bvol{287},  \pg{317--348}.

\bibitem[Hultmark \& Smits(2010)]{hultmark:2010a}
{\sc \au{Hultmark, M.} \& \au{Smits, A.~J.}} \yr{2010}  \at{Temperature
  corrections for constant temperature and constant current hot-wire
  anemometers}.  \jt{Meas. Sci. Technol.}  \bvol{21}, 105404.

\bibitem[Hutchins {\em et~al.\/}(2011)Hutchins, Monty, Ganapathisubramani, Ng
  \& Marusic]{hutchins:2011a}
{\sc \au{Hutchins, N.}, \au{Monty, J.~P.}, \au{Ganapathisubramani, B.}, \au{Ng,
  H. C.~H.} \& \au{Marusic, I.}} \yr{2011}  \at{Three-dimensional conditional
  structure of a high {R}eynolds number turbulent boundary layer}.  \jt{J.
  Fluid Mech.}  \bvol{673},  \pg{255--285}.

\bibitem[Hutchins {\em et~al.\/}(2009)Hutchins, Nickels, Marusic \&
  Chong]{hutchins:2009a}
{\sc \au{Hutchins, N.}, \au{Nickels, T.~B.}, \au{Marusic, I.} \& \au{Chong,
  M.~S.}} \yr{2009}  \at{Hot-wire spatial resolution issues in wall-bounded
  turbulence}.  \jt{J. Fluid Mech.}  \bvol{635},  \pg{103--136}.

\bibitem[Hwang {\em et~al.\/}(2009)Hwang, Bonness \& Hambric]{hwang:2009a}
{\sc \au{Hwang, Y.~F.}, \au{Bonness, W.~K.} \& \au{Hambric, S.~A.}} \yr{2009}
  \at{Comparison of semi-empirical models for turbulent boundary layer wall
  pressure spectra}.  \jt{J. Sound Vibr.}  \bvol{319},  \pg{199--217}.

\bibitem[Ingard(1953)]{ingard:1953a}
{\sc \au{Ingard, U.}} \yr{1953}  \at{On the theory and design of acoustic
  resonators}.  \jt{J. Acoust. Soc. Am.}  \bvol{25}~(6),  \pg{1037--1061}.

\bibitem[Jim\'{e}nez \& Hoyas(2008)]{jimenez:2008a}
{\sc \au{Jim\'{e}nez, J.} \& \au{Hoyas, S.}} \yr{2008}  \at{Turbulent
  fluctuations above the buffer layer of wall-bounded flows}.  \jt{J. Fluid
  Mech.}  \bvol{611},  \pg{215--236}.

\bibitem[Kasagi {\em et~al.\/}(2009)Kasagi, Suzuki \& Fukagata]{kasagi:2009a}
{\sc \au{Kasagi, N.}, \au{Suzuki, Y.} \& \au{Fukagata, K.}} \yr{2009}
  \at{Microelectromechanical systems--based feedback control of turbulence for
  skin friction reduction}.  \jt{Annu. Rev. Fluid Mech.}  \bvol{41},
  \pg{231--251}.

\bibitem[Klewicki {\em et~al.\/}(2008)Klewicki, Priyadarshana \&
  Metzger]{klewicki:2008a}
{\sc \au{Klewicki, J.~C.}, \au{Priyadarshana, P. J.~A.} \& \au{Metzger, M.~M.}}
  \yr{2008}  \at{Statistical structure of the fluctuating wall pressure and its
  in-plane gradients at high {R}eynolds number}.  \jt{J. Fluid Mech.}
  \bvol{609},  \pg{195--220}.

\bibitem[Lasagna {\em et~al.\/}(2013)Lasagna, Orazi \& Iuso]{lasagna:2013a}
{\sc \au{Lasagna, D.}, \au{Orazi, M.} \& \au{Iuso, G.}} \yr{2013}
  \at{Multi-time delay, multi-point linear stochastic estimation of a cavity
  shear layer velocity from wall-pressure measurements}.  \jt{Phys. Fluids}
  \bvol{25}, 017101.

\bibitem[Lee {\em et~al.\/}(1998)Lee, Kim \& Choi]{lee:1998a}
{\sc \au{Lee, C.}, \au{Kim, J.} \& \au{Choi, H.}} \yr{1998}  \at{Suboptimal
  control of turbulent channel flow for drag reduction}.  \jt{J. Fluid Mech.}
  \bvol{358},  \pg{245--258}.

\bibitem[Lee \& Moser(2015)]{lee:2015a}
{\sc \au{Lee, M.} \& \au{Moser, R.~D.}} \yr{2015}  \at{Direct numerical
  simulation of turbulent channel flow up to ${R}e_\tau = 5200$}.  \jt{J. Fluid
  Mech.}  \bvol{774},  \pg{395--415}.

\bibitem[Lee \& Moser(2019)]{lee:2019a}
{\sc \au{Lee, M.} \& \au{Moser, R.~D.}} \yr{2019}  \at{Spectral analysis of the
  budget equation in turbulent channel flows at high {R}eynolds number}.
  \jt{J. Fluid Mech.}  \bvol{860},  \pg{886--938}.

\bibitem[Liu \& Gayme(2020)]{liu:2020a}
{\sc \au{Liu, C.} \& \au{Gayme, D.~F.}} \yr{2020}  \at{An input--output based
  analysis of convective velocity in turbulent channels}.  \jt{J. Fluid Mech.}
  \bvol{888},  \pg{A32}.

\bibitem[Madhusudanan {\em et~al.\/}(2019)Madhusudanan, Illingworth \&
  Marusic]{madhusudanan:2019a}
{\sc \au{Madhusudanan, A.}, \au{Illingworth, S.~J.} \& \au{Marusic, I.}}
  \yr{2019}  \at{Coherent large-scale structures from the linearized
  {N}avier-{S}tokes equations}.  \jt{J. Fluid Mech.}  \bvol{873},
  \pg{89--109}.

\bibitem[Mathis {\em et~al.\/}(2009)Mathis, Hutchins \& Marusic]{mathis:2009a}
{\sc \au{Mathis, R.}, \au{Hutchins, N.} \& \au{Marusic, I.}} \yr{2009}
  \at{Large-scale amplitude modulation of the small-scale structures in
  turbulent boundary layers}.  \jt{J. Fluid Mech.}  \bvol{628},  \pg{311--337}.

\bibitem[{Merino-Mart\'{i}nez} {\em et~al.\/}(2020){Merino-Mart\'{i}nez},
  {Rubio Carpio}, Pereira, {van Herk}, Avallone, Ragni \&
  Kotsonis]{merino:2020a}
{\sc \au{{Merino-Mart\'{i}nez}, R.}, \au{{Rubio Carpio}, A.}, \au{Pereira,
  L.~Lima}, \au{{van Herk}, S.}, \au{Avallone, F.}, \au{Ragni, D.} \&
  \au{Kotsonis, M.}} \yr{2020}  \at{{Aeroacoustic design and characterization
  of the 3D-printed, open-jet, anechoic wind tunnel of Delft University of
  Technology}}.  \jt{Appl. Acoust.}  \bvol{170},  \pg{107504}.

\bibitem[Murray \& Ukeiley(2003)]{murray:2003a}
{\sc \au{Murray, N.~E.} \& \au{Ukeiley, L.~S.}} \yr{2003}  \at{Estimation of
  the flowfield from surface pressure measurements in an open cavity}.
  \jt{AIAA J.}  \bvol{41}~(5),  \pg{969--972}.

\bibitem[Murray \& Ukeiley(2004)]{murray:2004c}
{\sc \au{Murray, N.~E.} \& \au{Ukeiley, L.~S.}} \yr{2004} Low-dimensional
  estimation of cavity flow dynamics.  \bt{In {\em AIAA Paper 2004-681\/}}.

\bibitem[Naguib {\em et~al.\/}(1996)Naguib, Gravante \& Wark]{naguib:1996a}
{\sc \au{Naguib, A.~M.}, \au{Gravante, S.~P.} \& \au{Wark, C.~E.}} \yr{1996}
  \at{Extraction of turbulent wall-pressure time-series using an optimal
  filtering scheme}.  \jt{Exp. Fluids}  \bvol{22},  \pg{14--22}.

\bibitem[Naguib {\em et~al.\/}(2001)Naguib, Wark \&
  Juckenh\"{o}ofel]{naguib:2001a}
{\sc \au{Naguib, A.~M.}, \au{Wark, C.~E.} \& \au{Juckenh\"{o}ofel, O.}}
  \yr{2001}  \at{Stochastic estimation and flow sources associated with surface
  pressure events in a turbulent boundary layer}.  \jt{Phys. Fluids}
  \bvol{13}~(9),  \pg{2611--2626}.

\bibitem[Naka {\em et~al.\/}(2015)Naka, Stanislas, Foucaut, Coudert, Laval \&
  Obi]{naka:2015a}
{\sc \au{Naka, Y.}, \au{Stanislas, M.}, \au{Foucaut, {J.-M.}}, \au{Coudert,
  S.}, \au{Laval, {J.-P.}} \& \au{Obi, S.}} \yr{2015}  \at{Space--time
  pressure--velocity correlations in a turbulent boundary layer}.  \jt{J. Fluid
  Mech.}  \bvol{771},  \pg{624--675}.

\bibitem[Orszag(1971)]{orszag:1971a}
{\sc \au{Orszag, S.~A.}} \yr{1971}  \at{On the elimination of aliasing in
  finite-difference schemes by filtering high-wavenumber components}.  \jt{J.
  Atmos. Sci.}  \bvol{28}~(6),  \pg{1074--1074}.

\bibitem[Panton {\em et~al.\/}(2017)Panton, Lee \& Moser]{panton:2017a}
{\sc \au{Panton, R.~L.}, \au{Lee, M.} \& \au{Moser, R.~D.}} \yr{2017}
  \at{Correlation of pressure fluctuations in turbulent wall layers}.
  \jt{Phys. Rev. Fluids}  \bvol{2}, 094604.

\bibitem[Panton \& Miller(1975)]{panton:1975b}
{\sc \au{Panton, R.~L.} \& \au{Miller, J.~M.}} \yr{1975}  \at{Resonant
  frequencies of cylindrical {H}elmholtz resonators}.  \jt{J. Acoust. Soc. Am.}
   \bvol{57}~(6),  \pg{1533--1535}.

\bibitem[Perry \& Li(1990)]{perry:1990a}
{\sc \au{Perry, A.~E.} \& \au{Li, J.~D.}} \yr{1990}  \at{Experimental support
  for the attached-eddy hypothesis in zero-pressure-gradient turbulent boundary
  layers}.  \jt{J. Fluid Mech.}  \bvol{218},  \pg{405--438}.

\bibitem[Qiao {\em et~al.\/}(2018)Qiao, Wu \& Zhou]{qiao:2018a}
{\sc \au{Qiao, Z.~X.}, \au{Wu, Z.} \& \au{Zhou, Y.}} \yr{2018}  \at{Turbulent
  boundary layer manipulation under a proportional-derivative closed-loop
  scheme}.  \jt{Phys. Fluids}  \bvol{30}, 115101.

\bibitem[Rathnasingham \& Breuer(2003)]{rathnasingham:2003a}
{\sc \au{Rathnasingham, R.} \& \au{Breuer, K.~S.}} \yr{2003}  \at{Active
  control of turbulent boundary layers}.  \jt{J. Fluid Mech.}  \bvol{495},
  \pg{209--233}.

\bibitem[Renard \& Deck(2016)]{renard:2016a}
{\sc \au{Renard, N.} \& \au{Deck, S.}} \yr{2016}  \at{A theoretical
  decomposition of mean skin friction generation into physical phenomena across
  the boundary layer}.  \jt{J. Fluid Mech.}  \bvol{790},  \pg{339--367}.

\bibitem[Schewe(1983)]{schewe:1983a}
{\sc \au{Schewe, G.}} \yr{1983}  \at{On the structure and resolution of
  wall-pressure fluctuations associated with turbulent boundary-layer flow}.
  \jt{J. Fluid Mech.}  \bvol{134},  \pg{311--328}.

\bibitem[Schoppa \& Hussain(1998)]{schoppa:1998a}
{\sc \au{Schoppa, W.} \& \au{Hussain, F.}} \yr{1998}  \at{A large-scale control
  strategy for drag reduction in turbulent boundary layers}.  \jt{Phys. Fluids}
   \bvol{10}~(5),  \pg{1049--1051}.

\bibitem[Schultz \& Flack(2007)]{schultz:2007a}
{\sc \au{Schultz, M.~P.} \& \au{Flack, K.~A.}} \yr{2007}  \at{The rough-wall
  turbulent boundary layer from the hydraulically smooth to the fully rough
  regime}.  \jt{J. Fluid Mech.}  \bvol{580},  \pg{381--405}.

\bibitem[Smits {\em et~al.\/}(2021)Smits, Hultmark, Lee \&
  Pirozzoli]{smits:2021a}
{\sc \au{Smits, A.~J.}, \au{Hultmark, M.}, \au{Lee, M.} \& \au{Pirozzoli, S.}}
  \yr{2021}  \at{Reynolds stress scaling in the near-wall region of
  wall-bounded flows}.  \jt{J. Fluid Mech.}  \bvol{926},  \pg{A31}.

\bibitem[Smits {\em et~al.\/}(2011{\natexlab{{\em a\/}}})Smits, Mc{K}eon \&
  Marusic]{smits:2011a}
{\sc \au{Smits, A.~J.}, \au{Mc{K}eon, B.~J.} \& \au{Marusic, I.}}
  \yr{2011{\natexlab{{\em a\/}}}}  \at{High {R}eynolds number wall turbulence}.
   \jt{Annu. Rev. Fluid Mech.}  \bvol{43},  \pg{353--375}.

\bibitem[Smits {\em et~al.\/}(2011{\natexlab{{\em b\/}}})Smits, Monty,
  Hultmark, Bailey, Hutchins \& Marusic]{smits:2011HWa}
{\sc \au{Smits, A.~J.}, \au{Monty, J.}, \au{Hultmark, M.}, \au{Bailey, S.
  C.~C.}, \au{Hutchins, N.} \& \au{Marusic, I.}} \yr{2011{\natexlab{{\em
  b\/}}}}  \at{Spatial resolution correction for wall-bounded turbulence
  measurements}.  \jt{J. Fluid Mech.}  \bvol{676},  \pg{41--53}.

\bibitem[Thomas \& Bull(1983)]{thomas:1983a}
{\sc \au{Thomas, A. S.~W.} \& \au{Bull, M.~K.}} \yr{1983}  \at{On the role of
  wall-pressure fluctuations in deterministic motions in the turbulent boundary
  layer}.  \jt{J. Fluid Mech.}  \bvol{128},  \pg{283--322}.

\bibitem[Tinney {\em et~al.\/}(2006)Tinney, Coiffet, Delville, Glauser, Jordan
  \& Hall]{tinney:2006a}
{\sc \au{Tinney, C.~E.}, \au{Coiffet, F.}, \au{Delville, J.}, \au{Glauser,
  M.~N.}, \au{Jordan, P.} \& \au{Hall, A.~M.}} \yr{2006}  \at{On spectral
  linear stochastic estimation}.  \jt{Exp. Fluids}  \bvol{41}~(5),
  \pg{763--775}.

\bibitem[Tsuji {\em et~al.\/}(2007)Tsuji, Fransson, Alfredsson \&
  Johansson]{tsuji:2007a}
{\sc \au{Tsuji, Y.}, \au{Fransson, J. H.~M.}, \au{Alfredsson, P.~H.} \&
  \au{Johansson, A.~V.}} \yr{2007}  \at{Pressure statistics and their scaling
  in high-{R}eynolds-number turbulent boundary layers}.  \jt{J. Fluid Mech.}
  \bvol{585},  \pg{1--40}.

\bibitem[{Van Blitterswyk} \& Rocha(2017)]{blitterswyk:2017a}
{\sc \au{{Van Blitterswyk}, J.} \& \au{Rocha, J.}} \yr{2017}  \at{An
  experimental study of the wall-pressure fluctuations beneath low {R}eynolds
  number turbulent boundary layers}.  \jt{J. Acoust. Soc. Am.}  \bvol{141}~(2),
   \pg{1257--1268}.

\bibitem[Willmarth(1975)]{willmarth:1975a}
{\sc \au{Willmarth, W.~W.}} \yr{1975}  \at{Pressure fluctuations beneath
  turbulent boundary layers}.  \jt{Annu. Rev. Fluid Mech.}  \bvol{7},
  \pg{13--36}.

\bibitem[Yao {\em et~al.\/}(2018)Yao, Chen \& Hussain]{yao:2018a}
{\sc \au{Yao, J.}, \au{Chen, X.} \& \au{Hussain, F.}} \yr{2018}  \at{Drag
  control in wall-bounded turbulent flows via spanwise opposed wall-jet
  forcing}.  \jt{J. Fluid Mech.}  \bvol{852},  \pg{678--709}.

\bibitem[Zhang \& Chernyshenko(206)]{zhang:2016a}
{\sc \au{Zhang, C.} \& \au{Chernyshenko, S.~I.}} \yr{206}  \at{Quasisteady
  quasihomogeneous description of the scale interactions in near-wall
  turbulence}.  \jt{Phys. Rev. Fluids}  \bvol{1}, 014401.

\end{thebibliography}

\end{document}